\documentclass[twocolumn]{aastex631}
\usepackage{amsmath}

\newcommand{\teff}{\mbox{$T_{\rm eff}$}}

\newcommand{\logg}{\mbox{$\log g$}}

\newcommand{\apogee}{APOGEE}

\shortauthors{Behmard et al.}
\graphicspath{{./}{figures/}}
\begin{document}

\title{Elemental Abundances of \emph{Kepler} Objects of Interest in APOGEE DR17}

\author[0000-0003-0012-9093]{Aida Behmard}
\affiliation{Division of Geological and Planetary Sciences, California Institute of Technology, Pasadena, CA 91125, USA}

\author[0000-0001-5082-6693]{Melissa K. Ness}
\affiliation{Center for Computational Astrophysics, Flatiron Institute, 162 5th Avenue, New York, NY 10010, USA}
\affiliation{Department of Astronomy, Columbia University, Pupin Physics Laboratories, New York, NY 10027, USA}

\author[0000-0002-6993-0826]{Emily C. Cunningham}
\altaffiliation{NASA Hubble Fellow}
\affiliation{Department of Astronomy, Columbia University, Pupin Physics Laboratories, New York, NY 10027, USA}
\affiliation{Center for Computational Astrophysics, Flatiron Institute, 162 5th Avenue, New York, NY 10010, USA}
%\altaffiliation{Hubble Fellow}

\author[0000-0001-9907-7742]{Megan Bedell}
\affiliation{Center for Computational Astrophysics, Flatiron Institute, 162 5th Avenue, New York, NY 10010, USA}



\begin{abstract}
The elemental abundances of planet host stars can shed light on the conditions of planet forming environments. We test if individual abundances of 130 known/candidate planet hosts in \apogee\ are statistically different from those of a reference doppelg\"anger sample. The reference set comprises objects selected with the same \teff, \logg, [Fe/H], and [Mg/H] as each Kepler Object of Interest (KOI). We predict twelve individual abundances (X = C, N, O, Na, Al, Si, Ca, Ti, V, Cr, Mn, Ni) for the KOIs and their doppelg\"angers using a local linear model of these four parameters, training on ASPCAP abundance measurements for a sample of field stars with high fidelity (SNR $>$ 200) \apogee\ observations. We compare element prediction residuals (model$-$measurement) for the two samples and find them to be indistinguishable, given a high quality sample selection. We report median intrinsic dispersions of $\sim$0.038 dex and $\sim$0.041 dex, for the KOI and doppelg\"anger samples, respectively, for these elements. We conclude that the individual abundances at fixed \teff, \logg, [Fe/H], and [Mg/H] are unremarkable for known planet hosts. Our results establish an upper limit on the abundance precision required to uncover any chemical signatures of planet formation in planet host stars. 

%\textbf{(Should we also comment that we would need a reference sample that we know for sure does not include planet hosts? This might be hard... We may have to focus on easily detectable planet architectures e.g., hot Jupiters)}

%Here, we examine the compositions of 131 Kepler Objects of Interest (KOIs) observed with the Apache Point Observatory Galactic Evolution Experiment (APOGEE) by applying local linear models that predict a range of elemental abundances  given only fiducial tracers of supernovae production channels (Fe, Mg). The residuals between model-predicted and APOGEE Stellar Parameter and Chemical Abundances Pipeline (ASPCAP DR17) abundances encode information on processes beyond galactic chemical evolution, such as planet formation. The intrinsic dispersion of abundance residuals for the KOIs range from $\sim$0.019$-$0.17 dex, while those of a comparison sample of stars lacking planets range from $\sim$0.017$-$0.13 dex. The KOI residual intrinsic dispersion is particularly large for certain abundances, namely Na. However, the intrinsic dispersion differences between the KOIs and comparison sample disappear when we consider a subset of our sample selected with more stringent abundance error cuts ($<$0.07 dex). We conclude that larger samples of planet hosts with high precision abundances are needed to unveil the intricate connections between host star chemistry and planet formation. 
\end{abstract}



\section{Introduction} \label{sec:intro}
%\mkn{minor edits to first few paragraphs, broke up a few sentences to make them shorter and changed first sentence to not repeat first sentence of abstract}\\

The elemental abundances of planet host stars bear the fingerprint of the processes governing planet formation and evolution. For example, it is well established that stars hosting giant planets often have enhanced iron abundances ([Fe/H]; \citealt{gonzalez1997,heiter2003,santos2004,fischer2005}). This is typically regarded as evidence for the core accretion model of planet formation (e.g., \citealt{rice2003,ida2004,alibert2011,mordasini2012,maldonado2019}); the host star [Fe/H] can be considered a proxy for the solid surface density of protoplanetary disks. In this context, more solids translate to rapid growth of planetary cores that can reach a critical mass of $\sim$10 $M_{\oplus}$ before the disk gas dissipates. This enables accretion of a substantial gaseous envelope. The planet-[Fe/H] trend appears to weaken with decreasing planet mass/radius \citep{sousa2008,ghezzi2010,ghezzi2018,schlaufman2011,buchhave2012,buchhave2015,wang2015}, but becomes stronger with decreasing orbital period, particularly in the $P$ $\lesssim$ 10 days regime \citep{mulders2016,narang2018,petigura2018,wilson2018,sousa2019,ghezzi2021}. Thus, the distributions of planet masses, radii, and orbital periods are sculpted by the amount of available solids and therefore the host star metallicity and planet forming environment.

The connections between [Fe/H] and planet architectures are well-studied because there are many strong iron absorption lines in the spectra of solar-like stars, making it a relatively easy abundance to constrain. High precision abundances beyond iron are more challenging to measure, but can unveil more detailed relationships between host star chemistry and planet architectures. For example, \citet{adibekyan2012a} found that Fe-poor ($-$0.1 $<$ [Fe/H] $<$ 0.2 dex) hosts of small and giant planets exhibit enhanced [X/H] ratios for Mg, Al, Si, Sc, and Ti. The authors later examined a sample of even more Fe-depleted ($-$0.65 $<$ [Fe/H] $<$ $-$0.3 dex) stars that host small, rocky planets, and found strong enhancements in Ti \citep{adibekyan2012b}. Similarly, \citet{maldonado2018} found that Fe-poor host stars of cool Jupiters tend to be enhanced in $\alpha$-elements. These results suggest that other refractory elements can compensate for low iron content during planet building block formation (e.g., \citealt{bashi2019}).   

\begin{figure} 
\centering
    \vspace*{0.02in}
    \includegraphics[width=0.48\textwidth]{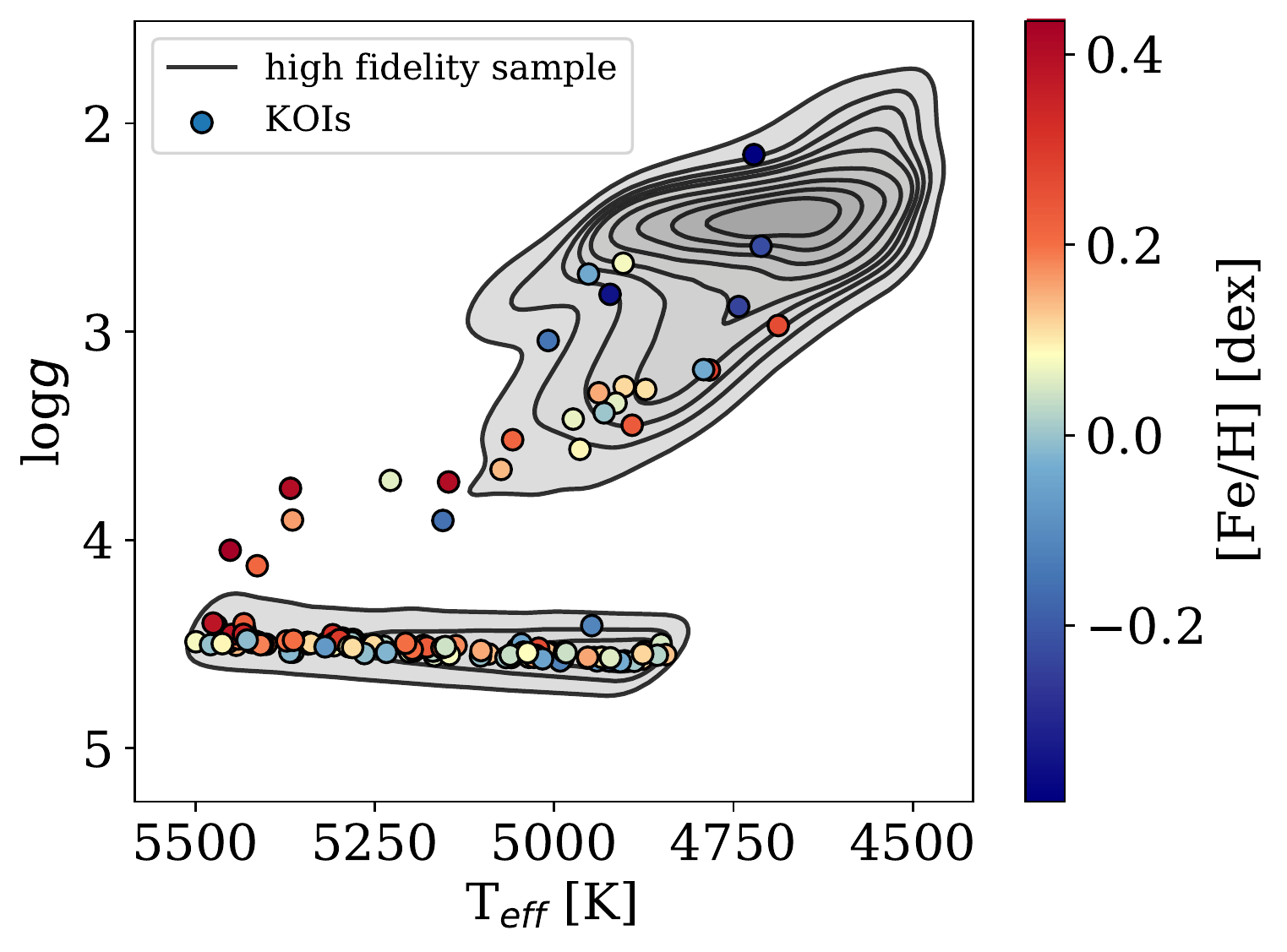}
    %\vspace*{-0.05in}
    \caption{The \logg \hspace{-0.1mm} vs. \teff \hspace{0.2mm} probability density functions for the $\sim$129,000 high fidelity APOGEE DR17 stars (gray). The contours represent areas encompassing 10\% of the cumulative probability mass. %\mkn{not sure understand how to read off fraction as explained as two groups - let's just chat in meeting what this sentence means?}. 
    The 130 non-`False Positive' KOIs included in the high fidelity sample with doppelg\"angers are shown as circles, with colors representing their [Fe/H] values.}
\label{fig:figure1}
\end{figure}

Abundances beyond iron can also place constraints on planet formation locations and interior compositions. For example, the stellar C/O ratio characterizes the H$_{2}$O, CO$_{2}$, and CO ice lines in protoplanetary disks, and can be used as a sensitive tracer of formation location when compared to the C/O ratio of planetary atmospheres \citep{oberg2011}; sub-stellar and super-stellar atmospheric C/O generally indicate planet formation within and beyond the H$_{2}$O ice line, respectively. The host star C/O ratio can also dictate if planetary compositions will be dominated by carbonates or silicates, with further ratios like Mg/Si determining the types of silicates in low C/O regimes (e.g., \citealt{brewer2017}).

Particular abundance patterns are also thought to be indicative of planet formation, as suggested by measured individual abundance trends with element condensation temperature ($T_{c}$). This is based on the premise that rocky planet-forming material more readily incorporates elements with high $T_{c}$ that reside in the solid phase throughout most of the disk. Conversely, low $T_{c}$ elements are more likely to remain in the gas phase. Planet compositions are thus characterized by larger abundances in order of increasing $T_{c}$. It follows that adding planetary material to host stars will create refractory enhancements in stellar photospheres and a positive abundance gradient with $T_{c}$. This could result from processes such as planet engulfment, or steady accretion of solids during Late Heavy Bombardment-like events. Depletion trends in order of $T_{c}$ could likewise result from an absence of planetary material in host star photospheres. This could result from solids getting locked up in rocky planets and subsequent accretion of dust-depleted gas onto the host star \citep{melendez2009}, or from gaps in protoplanetary disks created by forming giant planets that prevent host star accretion of refractory material \citep{booth2020}. Such trends with $T_{c}$ have been observed in the differential abundances of several binary systems \citep{ramirez2011,mack2014,tucci_maia2014,teske2015,ramirez2015,biazzo2015,saffe2016,teske2016,adibekyan2016,saffe2017,oh2018,tucci_maia2019,ramirez2019,nagar2020,galarza2021,jofre2021}, and in larger samples. For example, \citet{nibauer2021} analyzed 1700 solar analogs from the Apache Point Galactic Evolution Experiment (APOGEE), and found that 70$-$90\% of solar analogs appear depleted in refractory elements in order of $T_{c}$. Thus, there is ample evidence that abundance alteration via planet formation processes is common.

%\mkn{is it differential analyses of binary systems in all of this cited work - or just solar twins?} 

%\mkn{also think worth noting results of Nibauer 2019 here saying something like: Furthermore, larger survey analyses show that the majority of solar analogs look to be depleted in refractory elements and this is not a rare signature; less than $<$ 30 percent of stars do not show this trend (Nibauer 2019).}

\begin{figure} 
\centering
    \vspace*{0.02in}
    \includegraphics[width=0.5\textwidth]{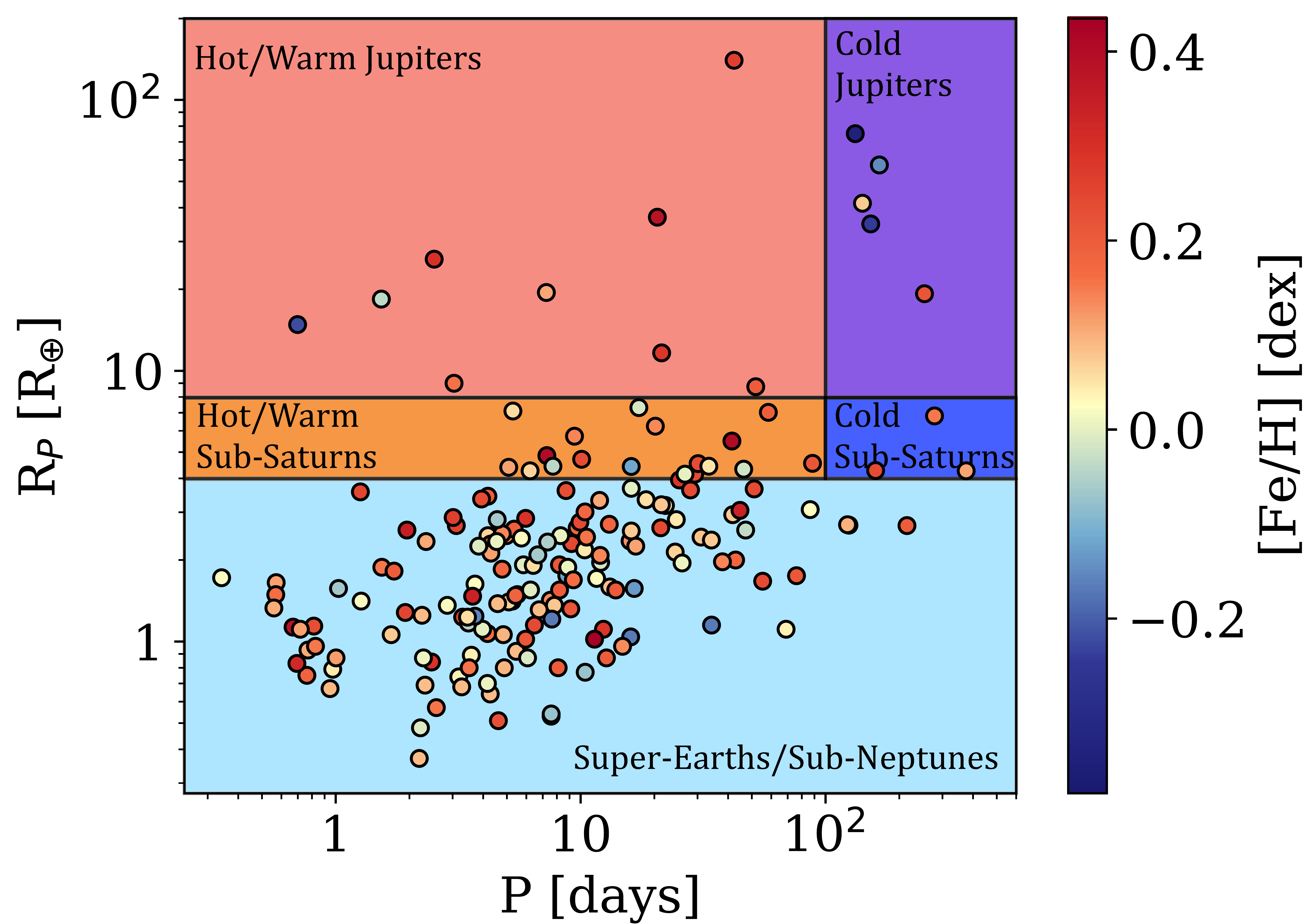}
    %\vspace*{0.05in}
    \caption{The radius vs. orbital period distribution for our sample of 130 KOI systems, with the ASPCAP-reported [Fe/H] of their host stars marked in color. Hot/warm Jupiters are defined as planets with $R$ $>$ 8 $R_{\oplus}$ and $P$ $<$ 100 days (red region), hot/warm sub-Saturns with 4 $R_{\oplus}$ $<$ $R$ $<$ 8 $R_{\oplus}$ and $P$ $<$100 days (orange region), cold Jupiters with $R$ $>$ 8 $R_{\oplus}$ and $P$ $>$ 100 days (purple region), cold sub-Saturns with 4 $R_{\oplus}$ $<$ $R$ $<$ 8 and $P$ $>$ 100 days (dark blue region), and super-Earths/sub-Neptunes with $R$ $<$ 4 $R_{\oplus}$ (light blue region).}
\label{fig:figure2}
\end{figure}

Stellar elemental abundances beyond iron are therefore important for understanding planet formation and evolution. Drawing connections between abundances and planet architectures require sufficiently large stellar samples to establish statistically significant correlations, as well as high precision ($\sim$0.01 dex uncertainties) abundance measurements \citep{melendez2009,ramirez2014,schuler2015}. Here, we utilize the latest APOGEE data release (DR17), which provides high-resolution spectra ($R$ $\approx$ 22,500) and derived parameters for $>$650,000 stars \citep{abdurrouf2022}. This enormous sample will boost abundance pattern statistics, making it possible to compromise on individual abundance precisions. The APOGEE DR17 parameters include individual abundances for 20 species, measured with the APOGEE Stellar Parameter and Chemical Abundances Pipeline (ASPCAP) pipeline \citep{garcia2016}. The full second generation APOGEE sample observed at the Apache Point Observatory (APOGEE-2N) contains 2098 stars also observed by Kepler, where 824 are confirmed planet hosts. This makes APOGEE DR17 an excellent sample for exploring connections between host star chemistry and planet formation. We describe our data selection further in Section \ref{sec:samples}.

%\mkn{We describe our data selection in Section 2. }

%However, constructing large high precision abundance samples requires high resolution, high signal-to-noise ratio (SNR) spectroscopy and automated spectral synthesis pipelines that can be time-intensive to employ. \mkn{In recent years such samples have become available.} 

%\mkn{note: some additional paragraphs and text below - please check} 

Our goal is to examine individual abundances in planet hosts in isolation of other parameters, such as evolutionary state and overall metallicity. We want to determine if the individual abundances are differentiable in any way from the underlying field population (where planet membership is unknown). To this end, we take the Kepler Objects of Interest (KOIs, defined as stars that host confirmed or candidate planets) observed in APOGEE, and construct a reference set of doppelg\"angers with identical \teff, \logg, [Fe/H], and [Mg/H] from the APOGEE field. Recent work has demonstrated that (Fe, Mg) alone capture the majority of abundance dimensionality for stars more metal-rich than [Fe/H] $> -1.0$ dex with surprising predictive power \citep{weinberg2019,griffith2021,weinberg2022,ness2022}. This is because these elements are fiducial tracers of two primary production sources, specifically core collapse supernovae and low mass stellar explosions. However, small individual abundance variations at fixed (Fe, Mg) may represent (at least in part) key additional information on stellar birth and evolutionary histories \citep{weinberg2022,ting2022,ness2022}. Individual abundances are inherited from birth and can be modified as a consequence of both internal (e.g., dredge up, \citealt{souto2019}) and external evolution (e.g., planet engulfment, \citealt{oh2018}). Therefore, abundance scatter in absence of (Fe, Mg) and evolutionary state contributions may encode abundance deviations from birth. Stars with planets may furthermore be born with different abundance distributions at fixed (Fe, Mg) compared to stars without.

%\mkn{question for Aida: are we more testing if planet hosts have different abundances from birth, or if they have changed from birth due to planet formation - my thinking is that planets change the birth abundance under the 'refractory' trend picture - but want to discuss. I think we could be clearer in the above paragraph as to what our question is - but want to chat first (i.e. can we say planet formation and accretion in the bracket above)?}

%\mkn{check additions okay?}\\
Rather than simply comparing the individual elemental abundance distributions of our KOI and doppelg\"anger samples, we use a four-parameter (\teff, \logg, [Fe/H], [Mg/H]) model to predict the individual abundances of both the doppelg\"anger and KOI sets. This approach enables a quantitative exploration of the relative predictive power these four parameters hold for abundances of KOI stars compared to those of the field population. It also allows us to examine element correlations if there are clear discrepancies between the KOI and doppelg\"anger samples.
%\mkn{add a `why' sentence here here: e.g. `This approach enables a quantitative exploration of the relative predictive power of the four-parameters for KOI stars compared to the field. This also enables a study of element correlations if there are clear discrepancies between samples'.} 
Our model is detailed in Section \ref{sec:model}. The stars we use to build our model are effectively drawn from the same underlying population as our doppelg\"angers in that none are confirmed/candidate planet hosts; we do not know their planet memberships. This enables us to examine how well we can predict each individual element while only considering our four predictors. We present the results of our abundance residual analysis in Section \ref{sec:results}, and discuss these results in the context of potential planet host star chemistry and planet formation connections in Section \ref{sec:discussion}.

\section{Data} \label{sec:samples}
We assemble a high fidelity sample of APOGEE DR17 stars with abundance measurements for twelve elements (X = C, N, O, Na, Al, Si, Ca, Ti, V, Cr, Mn, Ni). These abundances are determined by the ASPCAP pipeline \citep{garcia2016}, and are reported with respect to Fe. Because we are interested in abundance patterns resulting from planet formation with respect to hydrogen rather than enhancements with respect to iron, we convert the abundances as relative to hydrogen ([X/H] = [X/Fe] + [Fe/H]). We then apply the following quality cuts which leaves a sample of $\sim$129,000 stars (Figure \ref{fig:figure1}):

%\mkn{not quite sure this is the reason I think in the end it would work out the same but let's discuss? Propose to just delete the 'because Fe abundance..'} 

\vspace{5mm}
\noindent \teff \hspace{1mm}= 4500$-$5500 K
\vspace{0.8mm}
\newline \noindent \logg \hspace{1mm}$>$ 1.8
\vspace{0.8mm}
\newline \noindent SNR $>$ 80
\vspace{0.8mm}
\newline \noindent [X/Fe]$_{\textrm{error}}$ $<$ 0.1 dex
\vspace{0.8mm}
\newline \noindent Flag \texttt{ASPCAPFLAGS} not set to \texttt{STAR\_BAD}, \texttt{M\_H\_BAD}, \texttt{ALPHA\_M\_BAD}
\vspace{0.8mm}
\newline \noindent Flag \texttt{STARFLAGS} not set to \texttt{VERY\_BRIGHT\_NEIGHBOR}
\vspace{5mm}

We then cross-match the sample with a catalog of 2098 KOIs observed by APOGEE-2N (Caleb Ca\~nas, \emph{private correspondence}), resulting in 220 high fidelity KOI stars from APOGEE DR17. We cross-match the resulting sample with the final Kepler planet candidate catalog data release (DR25, \citealt{coughlin2017}) to obtain up-to-date candidate dispositions and planetary parameters. We subsequently remove all KOIs marked as `False Positive', which indicates that the detected signals are due to events other than exoplanet transits, e.g., eclipsing binaries. This cut leaves 128 confirmed planets and 56 planet candidates hosted by 131 APOGEE DR17 stars. As expected, the KOI-APOGEE DR17 sample is dominated by ``Kepler-like" architectures. That is, planets characterized as super-Earths or sub-Neptunes (e.g., \citealt{winn2015,yang2020}); $\sim$92\% of the KOIs fit this category by hosting confirmed/candidate planets with orbital periods and planet radii of $P$ $<$ 400 days and $R$ $<$ 4 $R_{\oplus}$, respectively (Figure \ref{fig:figure2}). 

\begin{figure*}[t]
    \centering
    \begin{minipage}{0.48\textwidth}
        \centering
        \includegraphics[width=0.99\textwidth]{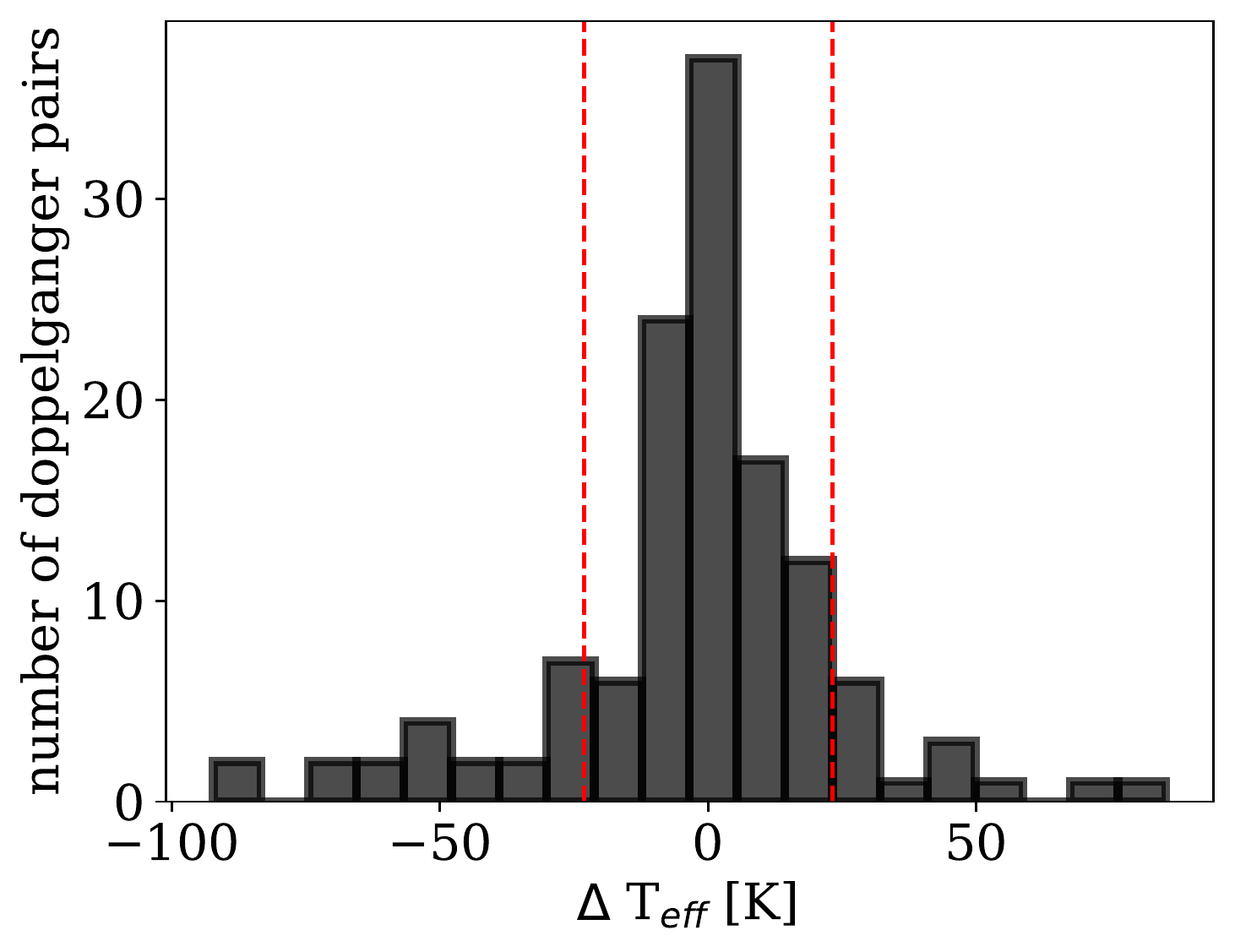} % first figure itself
    \end{minipage}\hfill
    \begin{minipage}{0.46\textwidth}
    \hspace{-4mm}
        \centering
        \includegraphics[width=0.99\textwidth]{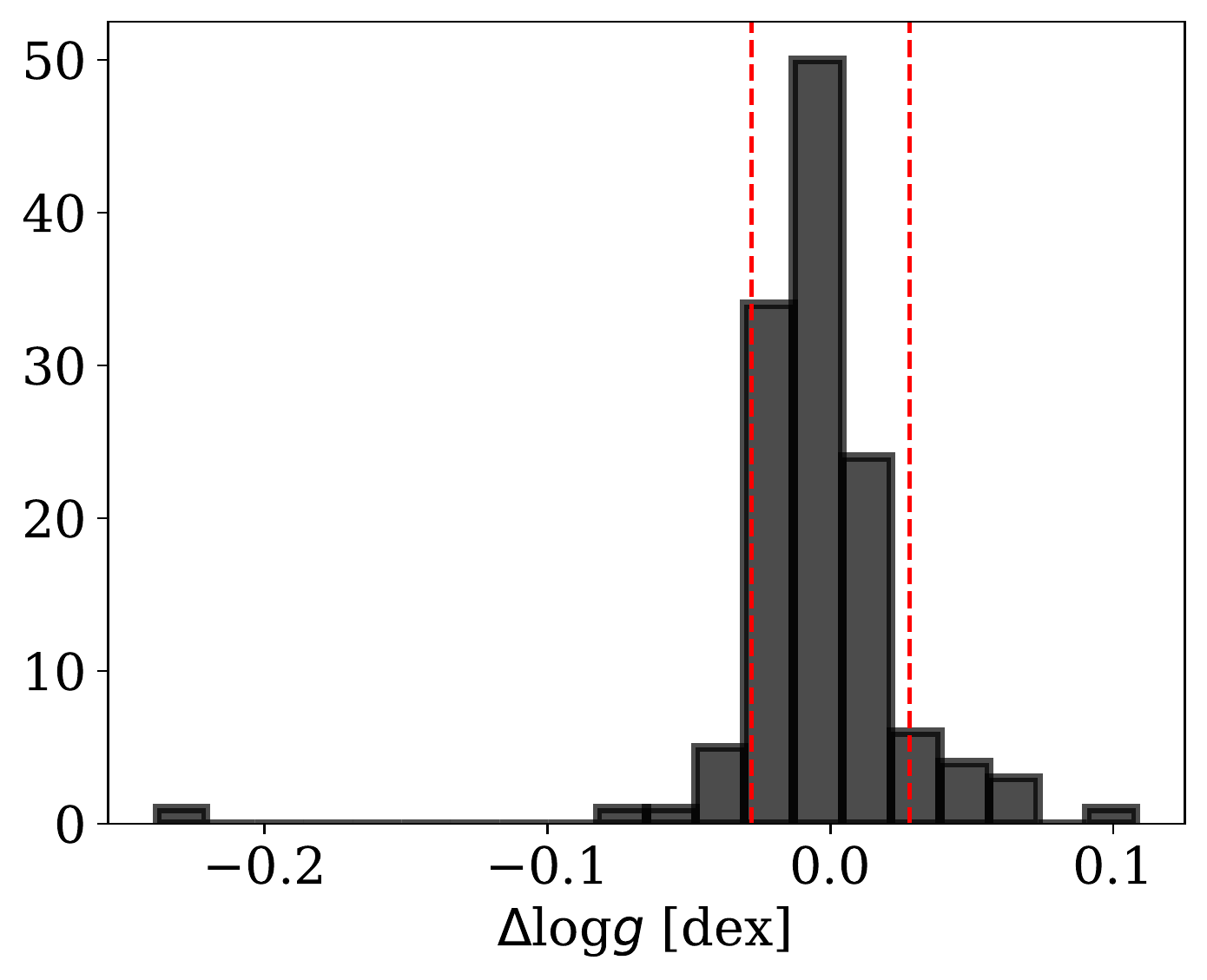} % second figure itself
    \end{minipage}
    \begin{minipage}{0.475\textwidth}
        \centering
        \includegraphics[width=0.99\textwidth]{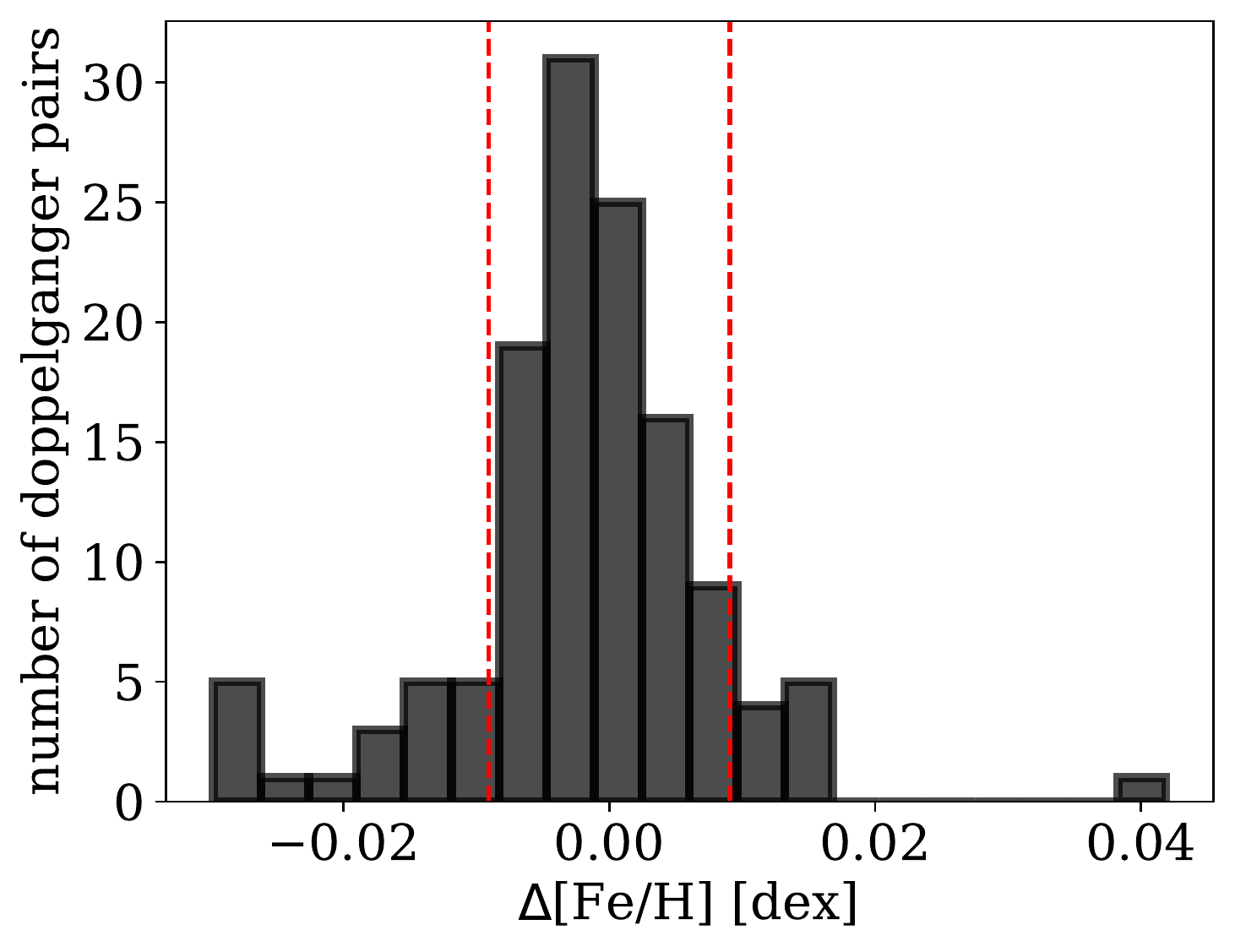} % first figure itself
    \end{minipage}\hfill
    \begin{minipage}{0.46\textwidth}
    %\vspace{-2mm}
        \centering
        \includegraphics[width=0.99\textwidth]{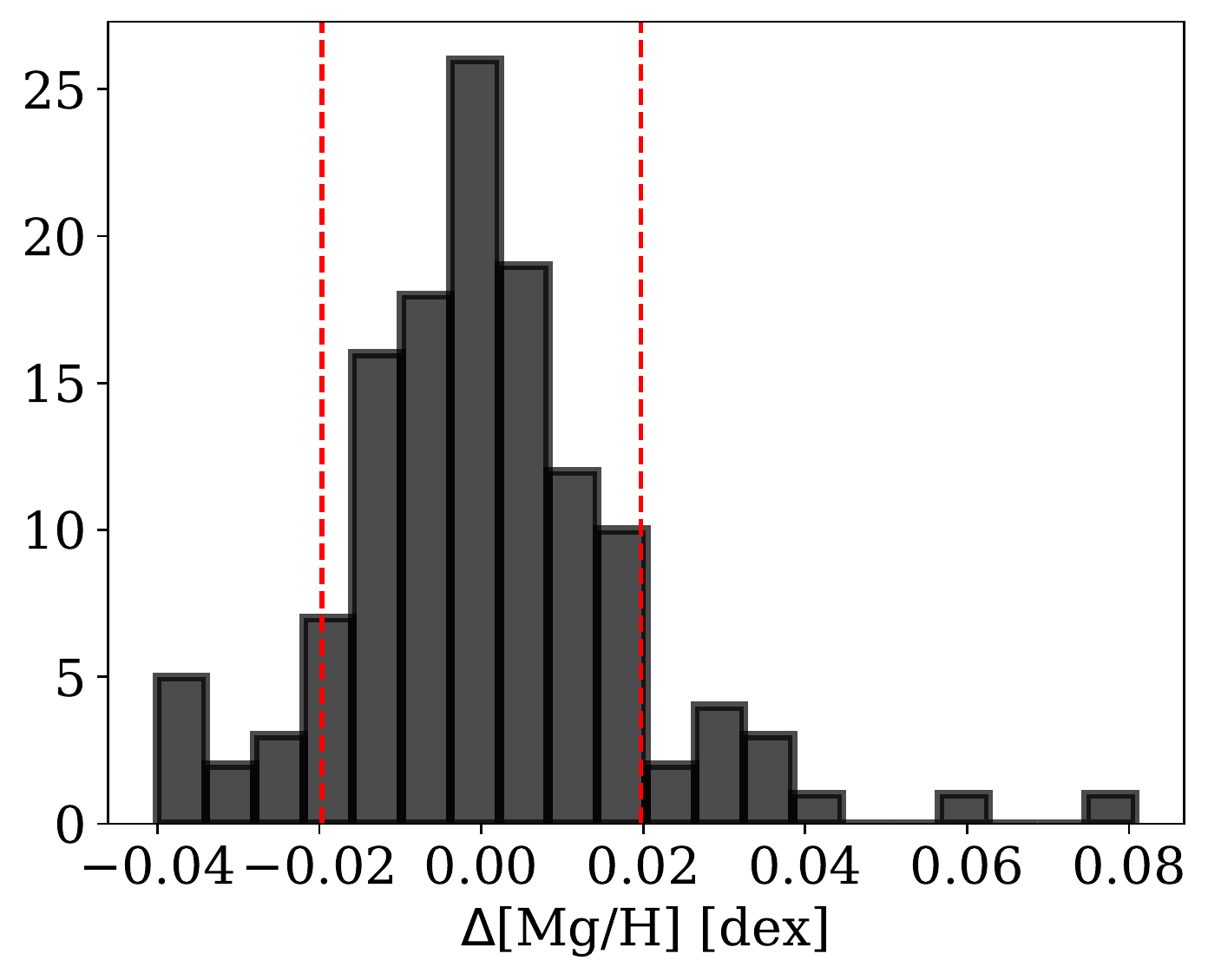} % second figure itself
    \end{minipage}
    \caption{$\Delta$\teff\ (upper left), $\Delta$\logg\ (upper right), $\Delta$[Fe/H] (lower left), and $\Delta$[Mg/H] (lower right) distributions for each KOI-doppelg\"anger pair. The average associated errors added in quadrature (for KOIs and doppelg\"angers) for each parameter are shown in the dashed red lines, and encompass the majority of the distributions ($\sim$77\%, $\sim$85\%, $\sim$78\%, and $\sim$81\% for the $\Delta$\teff, $\Delta$\logg, $\Delta$[Fe/H], and $\Delta$[Mg/H] distributions, respectively).}
\label{fig:figure3}
\end{figure*}

Next, we construct a set of doppelg\"anger stars to our KOI-APOGEE DR17 sample. The doppelg\"angers are drawn from the $\sim$129,000 high fidelity stars selected from the APOGEE field, and have unknown planet membership. We select doppelg\"angers by defining a similarity metric between two stars:

%\mkn{I think we need to say field doppelgangers and say planet membership is unknown - or perhaps I am wrong about this - is it true these are confirmed planet-less?} 

%\mkn{the denominators should have sqrt(2) for pairs as errors add in quadrature i.e. error$\_$t1$^$2 + error$\_$t2$^$2 - just read this is what these errors are further down - but perhaps state this before the equation as otherwise looks like the equation is wrong.)}

\begin{equation} \label{eq:equation1}
\begin{split}
D^{2} = \Big(\frac{\Delta T_{\textrm{eff}}} {\sqrt{\sigma_{T_{\textrm{eff}},1}^{2} + \sigma_{T_{\textrm{eff}},2}^{2}}}\Big)^{2}
+ \Big(\frac{\Delta \textrm{log}\hspace{0.5mm}g}
{\sqrt{\sigma_{\textrm{log}\hspace{0.5mm}g,1}^{2} + \sigma_{\textrm{log}\hspace{0.5mm}g,2}^{2}}}\Big)^{2} + \\
\Big(\frac{\Delta \textrm{[Fe/H]}}{\sqrt{\sigma_{\textrm{[Fe/H]},1}^{2} + \sigma_{\textrm{[Fe/H]},2}^{2}}}\Big)^{2} +  \\
\Big(\frac{\Delta\textrm{[Mg/H]}}{\sqrt{\sigma_{\textrm{[Mg/H]},1}^{2} + \sigma_{\textrm{[Mg/H]},2}^{2}}}\Big)^{2}
\end{split}
\end{equation}

\noindent which incorporates the relative \teff, \logg, [Fe/H], and [Mg/H] between the two stars, and their associated errors added in quadrature. These parameters are ideal for selecting doppelg\"angers because \teff\ and \logg\ describe the stellar evolutionary state, while [Fe/H] and [Mg/H] represent fiducial contributions from supernovae as mentioned earlier. Together, these parameters effectively create a four-dimensional reference frame to examine variance in individual elements. We select one doppelg\"anger per KOI, defined as the high fidelity non-KOI star drawn from the APOGEE field with the smallest $D^{2}$ metric value relative to that KOI, and SNR matching to within 20/pix. This SNR condition cannot be met for one KOI and any other star in the high fidelity sample, so we remove it and are left with a final KOI sample of 130 stars (Figure \ref{fig:figure1}, colored circles). The $\Delta$\teff, $\Delta$\logg, $\Delta$[Fe/H], and $\Delta$[Mg/H] distributions for all KOI-doppelg\"anger pairs are provided in Figure \ref{fig:figure3}. These distributions are centered on zero, which indicates that there are no systematic biases. The average associated errors added in quadrature (for KOIs and doppelg\"angers) for each parameter across all pairs are marked in the dashed red lines, which contain $\sim$77\%, $\sim$85\%, $\sim$78\%, and $\sim$81\% of the $\Delta$\teff, $\Delta$\logg, $\Delta$[Fe/H], and $\Delta$[Mg/H] distributions, respectively. Thus, the differences in parameters between KOIs and their respective doppelg\"angers are largely contained within their typical errors.

%\mkn{smallest D-squared could still be high in principle and mean we get doppelgangers that are larger than errors so not true doppelgangers if sparsely populated anywhere. I would cut off at D-squared $<$ 6? Could you include as figures a histogram of the mean $\Delta_$T, $\Delta_$logg, $\Delta_$Fe/H and $\Delta_$Mg with the denominator marked in dashed lines to show they are all true doppelgangers? This is a good figure for this section.}

We constructed another doppelg\"anger sample also selected on $K$-band extinction $A_{k}$ as provided by the \texttt{AK\_TARG} column in APOGEE DR17. The similarity metric is modified to include the $A_{k}$ term as follows:

\begin{equation} \label{eq:equation1}
\begin{split}
D_{A_{k}}^{2} = \Big(\frac{\Delta T_{\textrm{eff}}} {\sqrt{\sigma_{T_{\textrm{eff}},1}^{2} + \sigma_{T_{\textrm{eff}},2}^{2}}}\Big)^{2}
+ \Big(\frac{\Delta \textrm{log}\hspace{0.5mm}g}
{\sqrt{\sigma_{\textrm{log}\hspace{0.5mm}g,1}^{2} + \sigma_{\textrm{log}\hspace{0.5mm}g,2}^{2}}}\Big)^{2} + \\
\Big(\frac{\Delta \textrm{[Fe/H]}}{\sqrt{\sigma_{\textrm{[Fe/H]},1}^{2} + \sigma_{\textrm{[Fe/H]},2}^{2}}}\Big)^{2} + 
\Big(\frac{\Delta\textrm{[Mg/H]}}{\sqrt{\sigma_{\textrm{[Mg/H]},1}^{2} + \sigma_{\textrm{[Mg/H]},2}^{2}}}\Big)^{2}
+ \\
\Big(\frac{\Delta A_{k} }{\sqrt{\sigma_{A_{k},1}^{2}+\sigma_{A_{k},2}^{2}}}\Big)^{2}
\end{split}
\end{equation}

The $K$-band extinction characterizes the strength of absorption features in the optical and near-infrared wavelength range, e.g., diffuse interstellar bands (DIBs). These spectral features probe dusty regions of the interstellar medium (ISM), which is valuable from a planet formation perspective as planet occurrence is enhanced in metal-rich environments. We thus include this additional doppelg\"anger sample criterion for conducting a stricter test of similarity by also considering the line-of-sight ISM.

\section{Regression Model} \label{sec:model}
We construct a local linear model for each KOI to determine how well we can predict abundances from our four parameters of interest (\teff, \logg, [Fe/H], and [Mg/H]). We note that including the evolutionary state parameters accounts for any systematic changes in abundance with \teff\ or \logg. The local linear models employ simple linear regression \citep{hastie2001}, where each individual model is constructed from a training set specific to that KOI drawn from the high fidelity APOGEE DR17 sample of $\sim$129,000 stars. The training sets are selected by defining a region around each KOI in parameter space (e.g., \citealt{sayeed2021,ness2022}). We outline the steps of our approach below, where the parameters selected as predictors are $\vec{Y}$ = (\teff, \logg, [Fe/H], [Mg/H]), and the twelve predicted abundances are X = (C, N, O, Na, Al, Si, Ca, Ti, V, Cr, Mn, Ni):

\begin{enumerate}

    \item We standardize the parameters used as predictors across the entire high fidelity sample. This is done for each star by subtracting the mean and dividing by the standard deviation: y = ($Y$ - $\overline{Y}$)/$\sigma_{Y}$.
    
    \item For each star in our high fidelity sample, we identify the nearest $k$ neighbors in predictor parameter space according to the Euclidean distance. We carried this out with the \texttt{scikit-learn} (\texttt{sklearn}) package implemented in Python (specifically \texttt{sklearn.neighbors.KDTree}).
    
    \item We construct a local model for each KOI from the $k$ nearest non-KOI neighbors. We select $k$ = 100 but note that the model appears insensitive to the choice of $k$; $k$ = 50$-$300 produces comparable results \citep{ness2022}.
    
    \item We use linear regression, again applied via the \texttt{sklearn} package, to train each local linear model. This modeling step elucidates the relationships between the predictor parameters $\vec{Y}$ = (\teff, \logg, [Fe/H], [Mg/H]), and each of the twelve abundances [X/H], separately. Each local model includes five coefficients constrained from linear regression, corresponding to the intercept and one for each predictor parameter. 
    
    \item We predict a new set of twelve abundances for each KOI from their individual local linear models. The predicted [X/H] can be compared with the measured [X/H] from ASPCAP.
    
    \item We carry out this procedure for (i) our KOI-APOGEE DR17 sample and (ii) our corresponding doppelg\"anger samples. The result is a set of local linear models with one model per star for every star in each sample. We subsequently use these models for abundance prediction.

\end{enumerate}

\begin{figure*}[h]
    \centering
    \begin{minipage}{0.99\textwidth}
        \centering
        \includegraphics[width=0.99\textwidth]{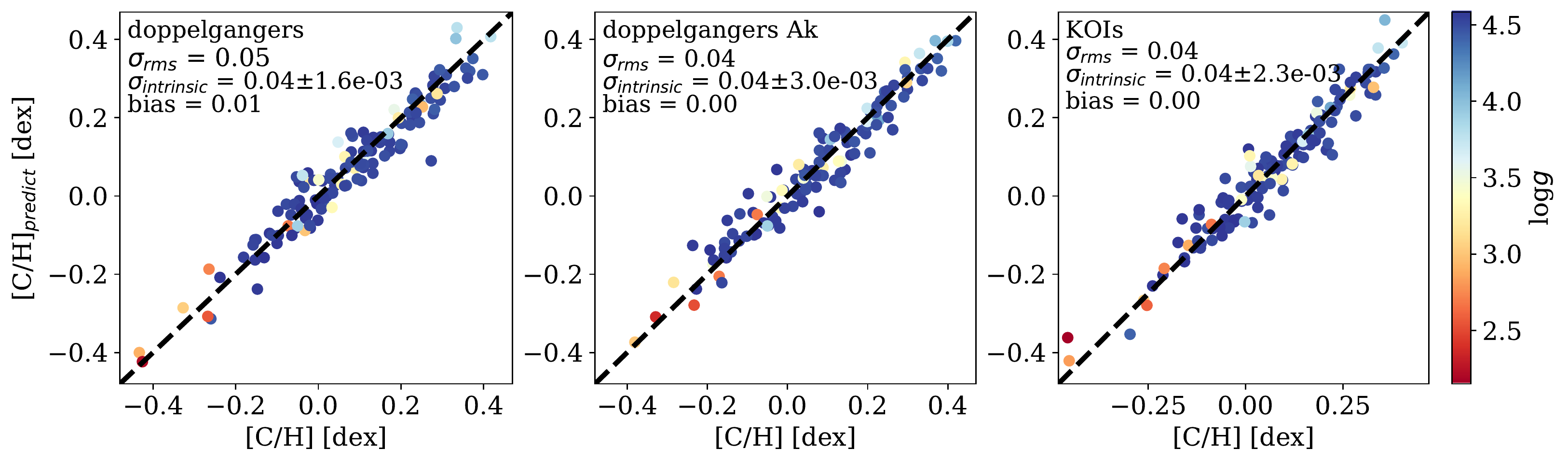} % first figure itself
    \end{minipage}\hfill
    \begin{minipage}{0.99\textwidth}
        \centering
        \includegraphics[width=0.99\textwidth]{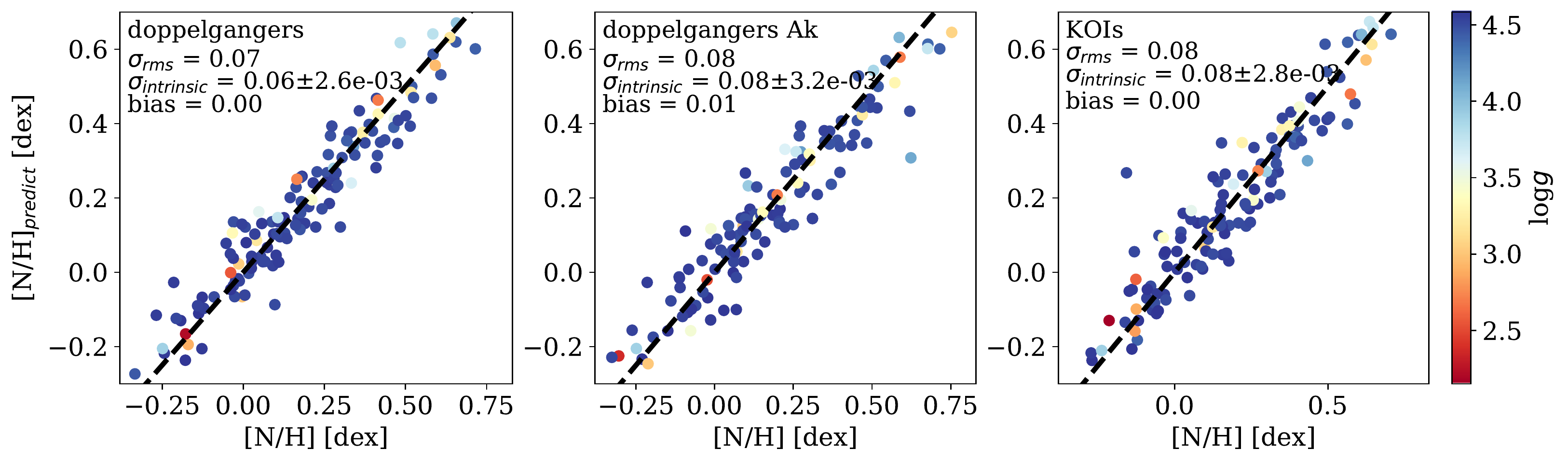} % second figure itself
    \end{minipage}\hfill
    \begin{minipage}{0.99\textwidth}
        \centering
        \includegraphics[width=0.99\textwidth]{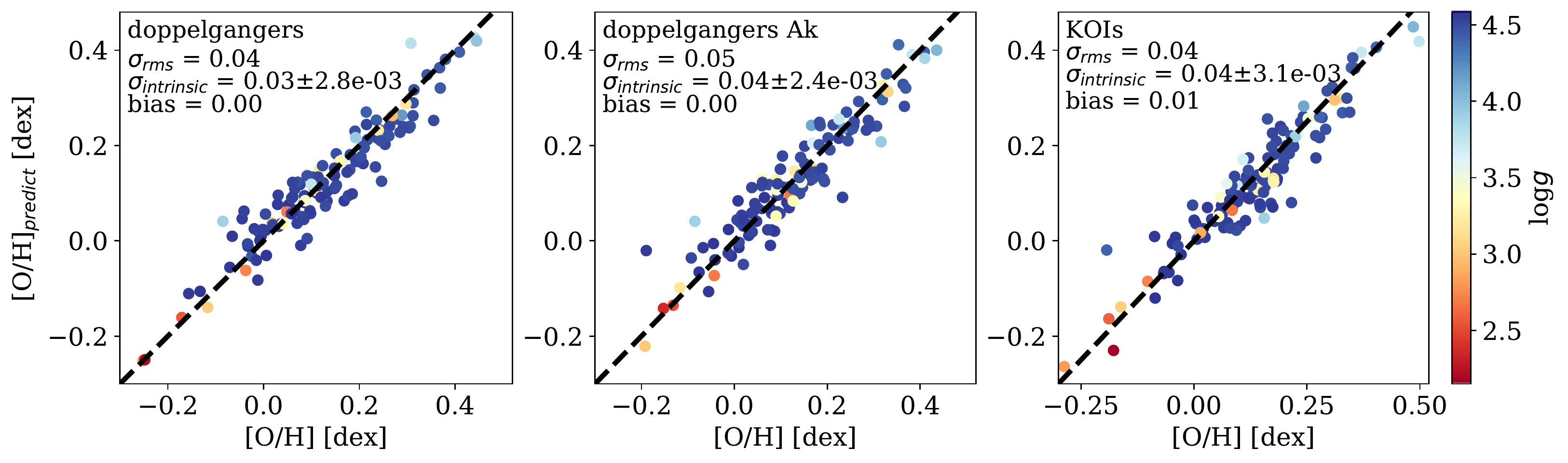} % second figure itself
    \end{minipage}\hfill
    \begin{minipage}{0.99\textwidth}
        \centering
        \includegraphics[width=0.99\textwidth]{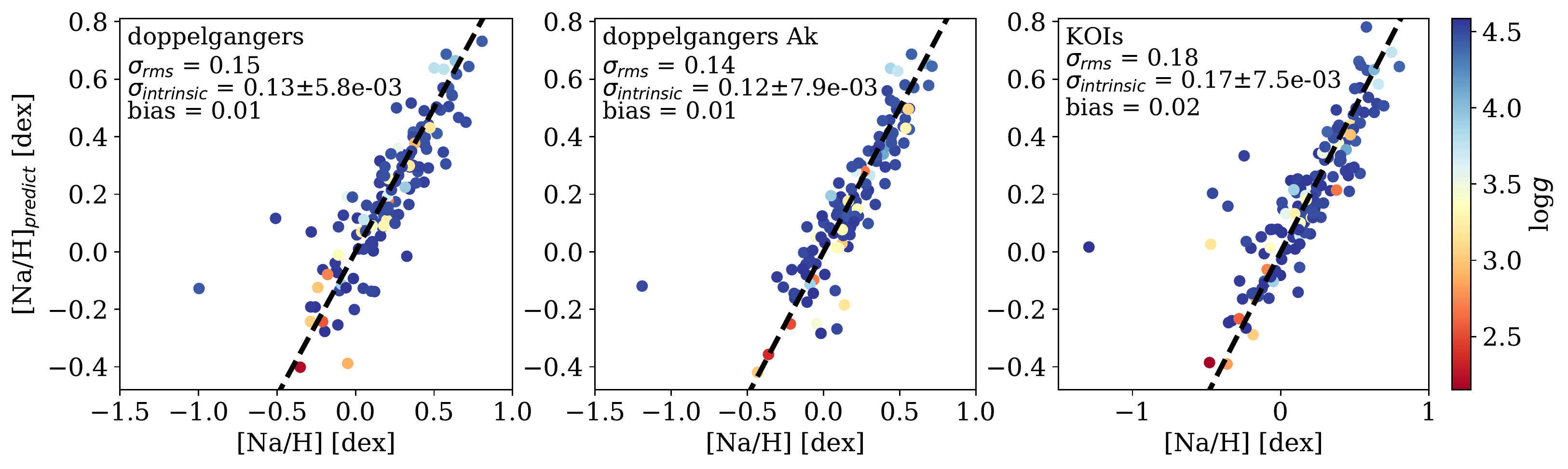} % second figure itself
    \end{minipage}
    \end{figure*}
\newpage

\begin{figure*}[h]
    \centering
    \begin{minipage}{0.99\textwidth}
        \centering
        \includegraphics[width=0.99\textwidth]{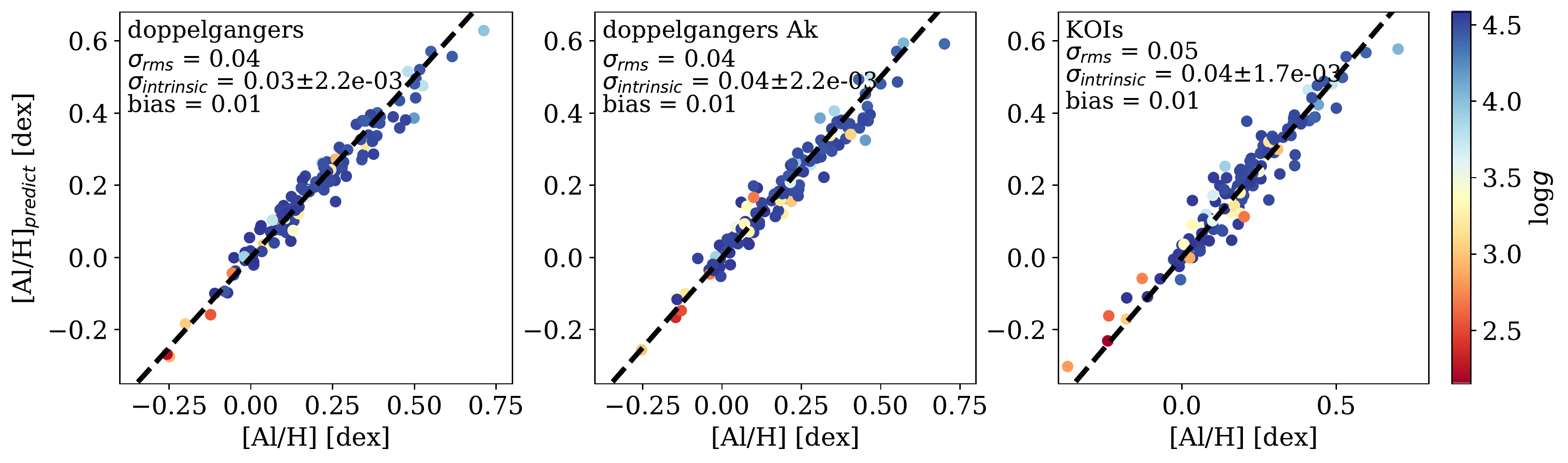} % second figure itself
    \end{minipage}\hfill
    \begin{minipage}{0.99\textwidth}
        \centering
        \includegraphics[width=0.99\textwidth]{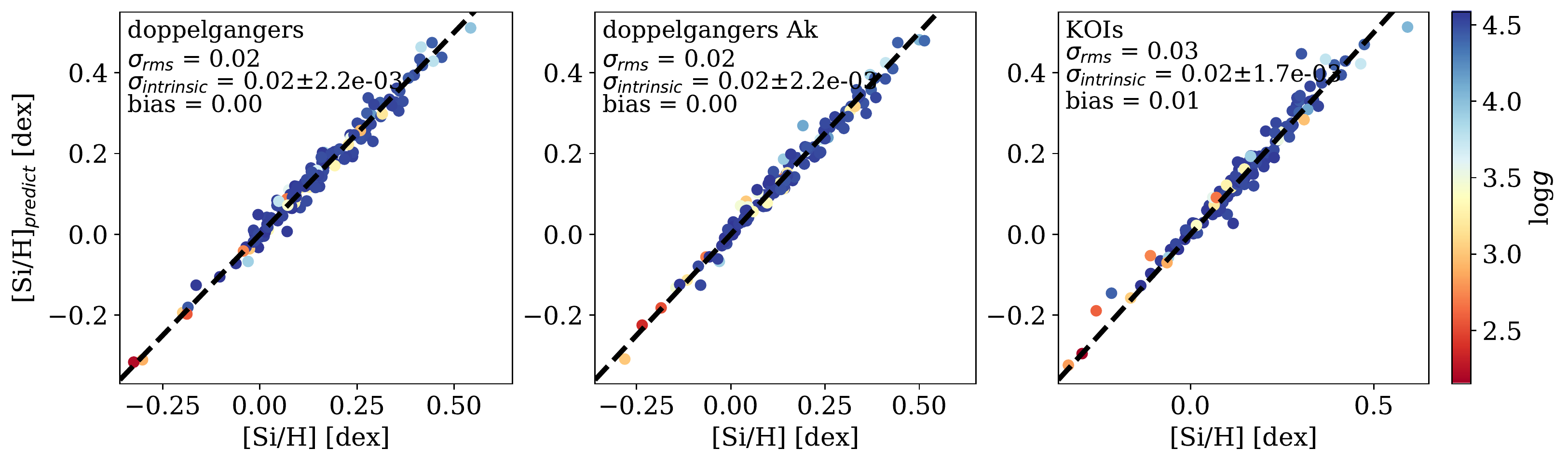} % second figure itself
    \end{minipage}\hfill
    \begin{minipage}{0.95\textwidth}
        \centering
        \includegraphics[width=0.99\textwidth]{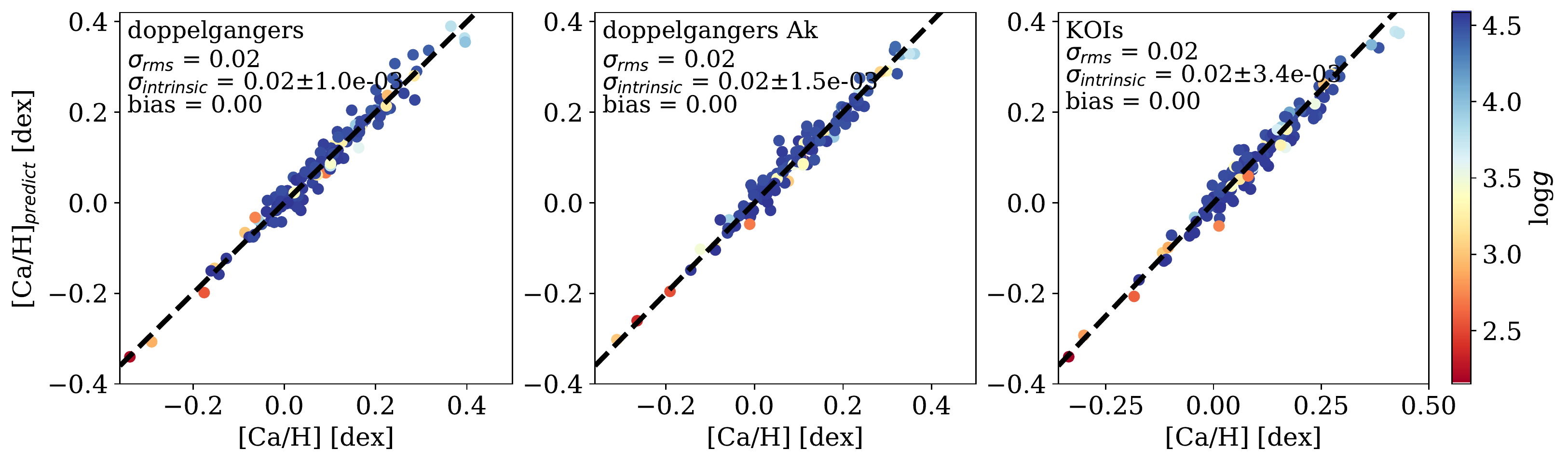} % second figure itself
    \end{minipage}\hfill
    \begin{minipage}{0.95\textwidth}
        \centering
        \includegraphics[width=0.99\textwidth]{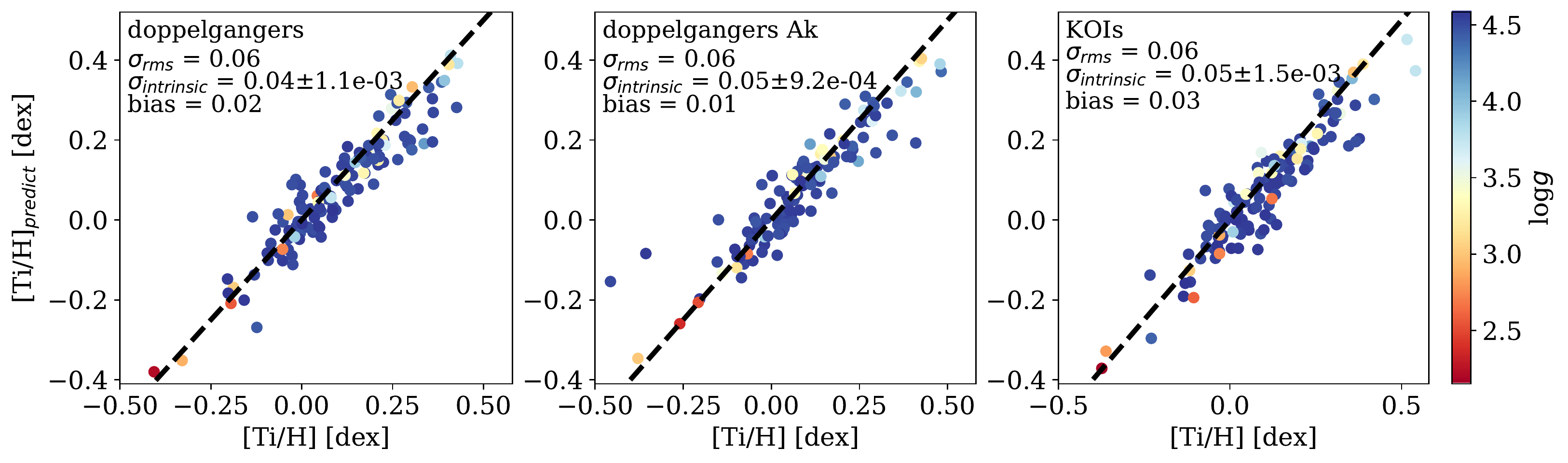} % second figure itself
    \end{minipage}
    \end{figure*}
\newpage

\begin{figure*}[h]
    \centering
    \begin{minipage}{0.99\textwidth}
        \centering
        \includegraphics[width=0.99\textwidth]{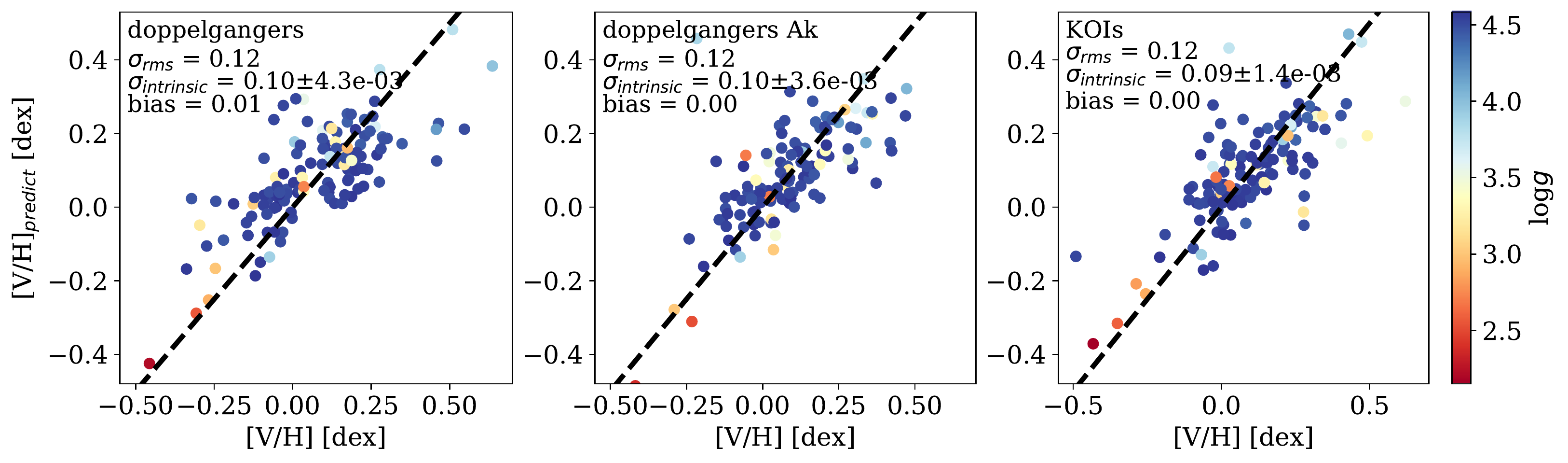} % first figure itself
    \end{minipage}\hfill
    \begin{minipage}{0.99\textwidth}
        \centering
        \includegraphics[width=0.99\textwidth]{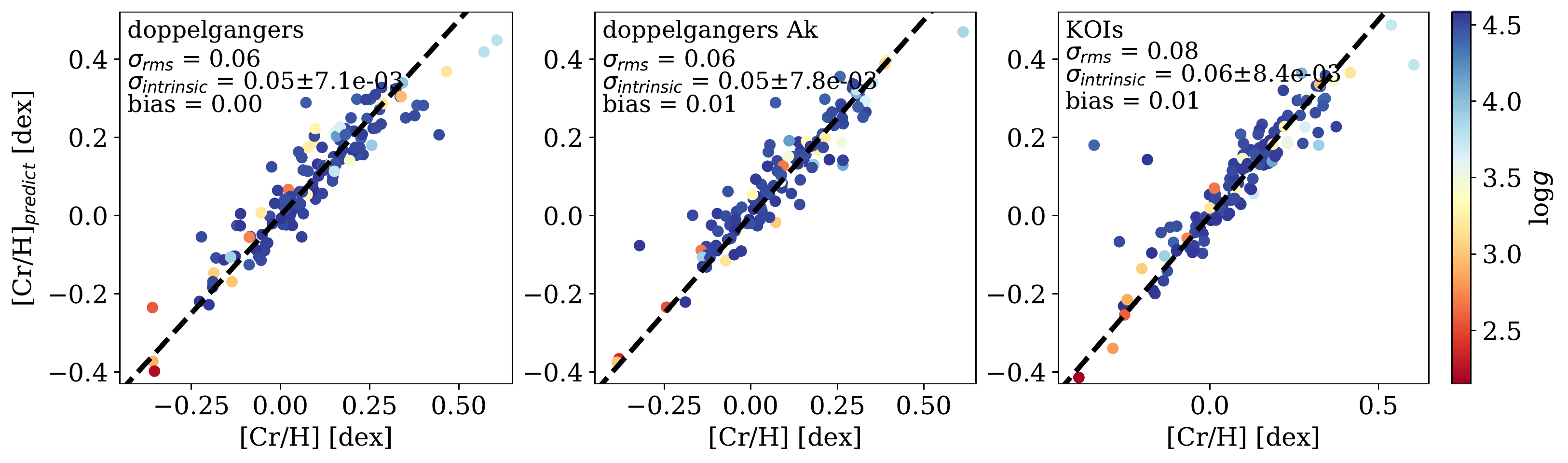} % second figure itself
    \end{minipage}\hfill
    \begin{minipage}{0.99\textwidth}
        \centering
        \includegraphics[width=0.99\textwidth]{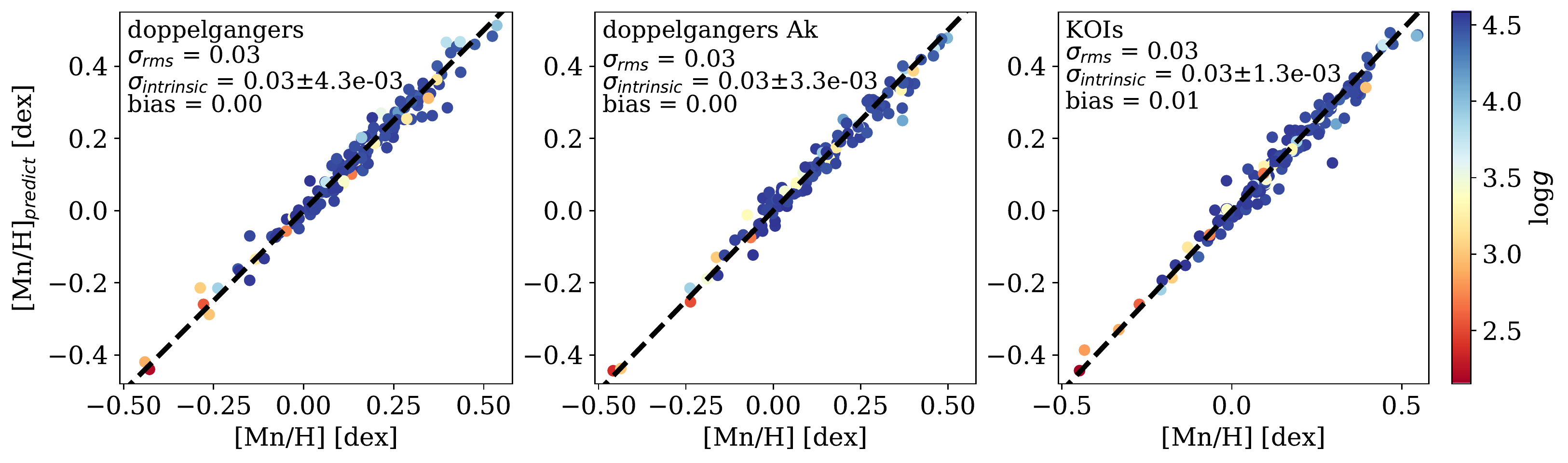} % second figure itself
    \end{minipage}\hfill
    \begin{minipage}{0.99\textwidth}
        \centering
        \includegraphics[width=0.99\textwidth]{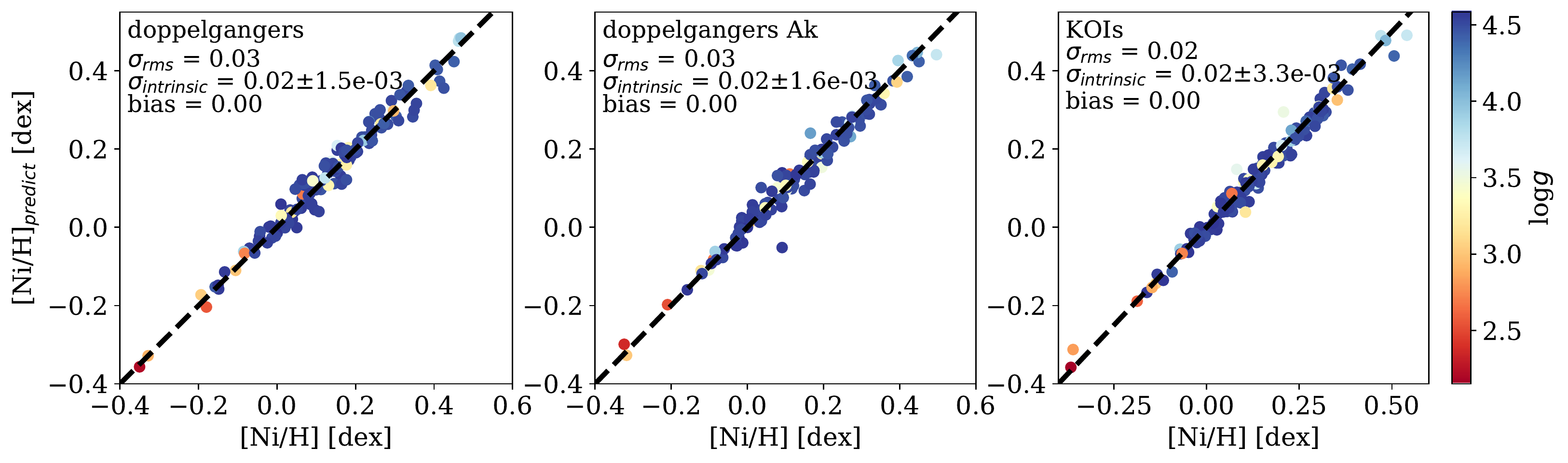} % second figure itself 
    \end{minipage}
    \caption{Local linear model-predicted vs. ASPCAP abundances of the doppelg\"anger (left), doppelg\"anger $A_{k}$ (middle), and KOI (right) samples for the twelve elements considered. The points are colored by log$g$. The rms difference between the ASPCAP and predicted abundances, intrinsic dispersions, and bias measurements are provided in the upper left corners of each panel. A dashed 1-to-1 line is plotted in all panels for comparison.}
\label{fig:figure4}
\end{figure*}

\section{Results} \label{sec:results}
\subsection{Local Linear Model Predictions} \label{sec:model_results}
Our local linear model-predicted abundances are plotted against ASPCAP abundances for the twelve considered elements in Figure \ref{fig:figure4}. The doppelg\"anger sample, doppelg\"anger sample also selected on $A_{k}$, and KOI sample are shown in the panels from left to right. We calculated the intrinsic dispersion of the abundance predictions from the root mean square (rms) difference between the model$-$measurement abundances, and the mean ASPCAP abundance uncertainty (which can be assumed as the same for each star): $\sigma_{\textrm{intrinsic}}$ = $\sqrt{\sigma^{2}_{\textrm{rms}} - \sigma^{2}_{\textrm{measurement}}}$. For the KOIs, the intrinsic dispersion values across all elements range from $\sigma_{\textrm{intrinsic}}$ = 0.019$-$0.167 dex, with Na and Ca exhibiting the highest and lowest values, respectively. If we group the abundances into light (C, N, O, Na, Al, V), alpha (Si, Ca), and iron-peak (Ti, Cr, Mn, Ni) element groups, the median intrinsic dispersion values are 0.060 dex, 0.021 dex, and 0.040 dex, respectively. For the doppelg\"angers, the intrinsic dispersion values range from 0.017$-$0.128 dex, with Na also exhibiting the highest value, but Si exhibiting the lowest value. The light, alpha, and iron-peak element groups have median intrinsic dispersion values of 0.053 dex, 0.017 dex, and 0.036 dex, respectively. We calculate the error on intrinsic dispersion $\sigma_{\textrm{intrinsic}}$ by sampling all abundances from their error distributions 20 times, running the local linear models, then taking the scatter of the resulting $\sigma_{\textrm{intrinsic}}$ values as the error on $\sigma_{\textrm{intrinsic}}$. We also calculate the abundance prediction bias as the difference of the mean predicted and ASPCAP abundances for each element.  

%\mkn{I think we should summarise the intrinsic dispersions as this is a result of the paper - could just list the range and smallest and largest elements, or group elements by family (light, alpha, iron-peak) and quite median results.}

\begin{figure*} 
\centering
    %\vspace*{0.02in}
    \includegraphics[width=0.999\textwidth]{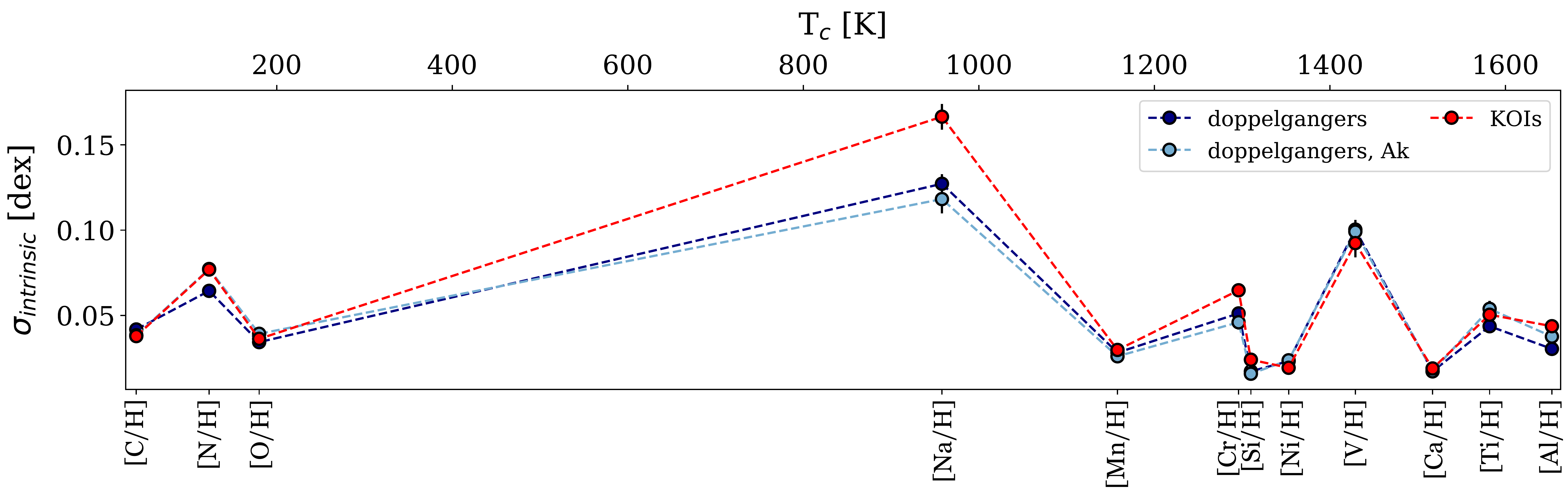}
    %\vspace*{-0.05in}
    \caption{Intrinsic dispersion measurements $\sigma_{\textrm{intrinsic}}$ for all $T_{c}$ ordered abundances, for the KOI sample (red), doppelg\"anger sample (navy), and $A_{k}$ doppelg\"anger sample (light blue). There are no apparent trends with $T_{c}$, but the KOI [Na/H] prediction intrinsic dispersion is noticeably larger than those of the doppelg\"anger samples.}
\label{fig:figure5}
\end{figure*}

The bias is approximately zero for the doppelg\"anger and KOI samples across all twelve abundances, which indicates that our predicted abundances are unbiased. The rms difference and intrinsic dispersion measurements are approximately equal across all abundances except for Na; for the KOIs, $\sigma_{\textrm{intrinsic}}$ $\approx$ 0.17 dex, while for the doppelg\"anger and $A_{k}$ doppelg\"anger samples, $\sigma_{\textrm{intrinsic}}$ $\approx$ 0.13 dex and 0.12 dex, respectively. We plot the intrinsic dispersion measurements for all abundances and samples in order of their condensation temperature $T_{c}$ in Figure \ref{fig:figure5}. There is no apparent $T_{c}$ trend, and the largest difference between the intrinsic dispersion values of the KOI and doppelg\"anger samples for [Na/H] is apparent.

\begin{figure} 
\centering
    \vspace*{0.02in}
    \includegraphics[width=0.5\textwidth]{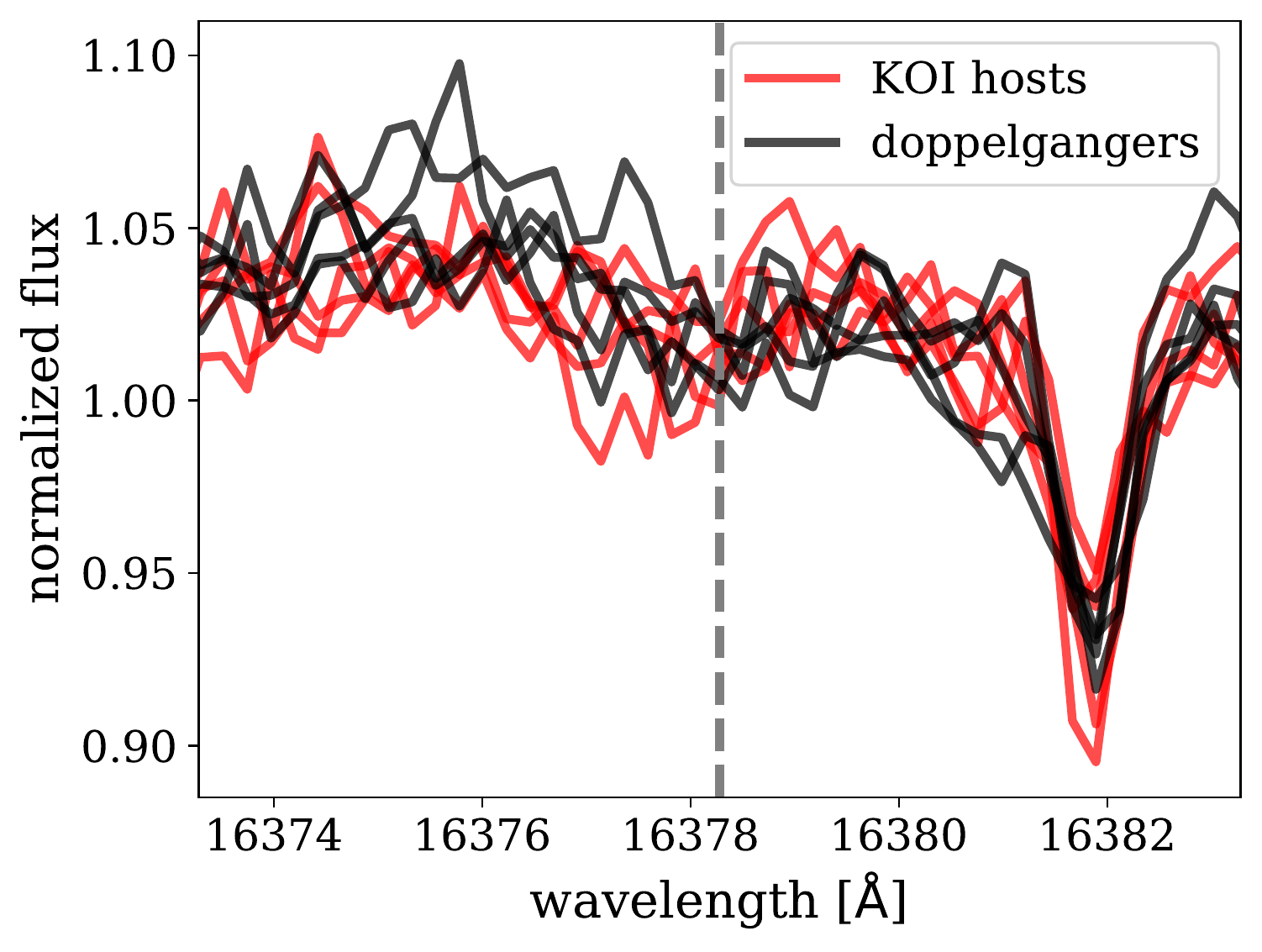}
    %\vspace*{-0.05in}
    \caption{Spectra of the five KOIs that appear to be outliers in predicted and ASPCAP [Na/H] space (red), zoomed into the region with Na spectral features used to derive [Na/H] abundances with the ASPCAP pipeline (window center at 16378.276 \AA, \citealt{feeney2021}, marked by the dashed line). The spectra of the corresponding outlier doppelg\"angers are shown in black.} .   %\mkn{for this figure we should put a dashed line in the center of the window and shade the window to clarify which is the Na feature})}}
\label{fig:figure6}
\end{figure}

The large intrinsic dispersion value for predicted [Na/H] exhibited by the KOI sample appears to be driven by five outlier stars that lie anomalously far from the 1-to-1 trend (Figure \ref{fig:figure4}, [Na/H] KOI panel). We examine the spectra of these outlier stars near the Na spectral features used to derive [Na/H] in the ASPCAP pipeline (window centered on 16378.276 \AA, \citealt{feeney2021}), shown in Figure \ref{fig:figure6}. There are no obvious differences between the Na features of the KOIs and their corresponding doppelg\"angers. Thus, we propose that the differences in intrinsic dispersion values between the KOIs and doppelg\"angers are due to poorly measured Na values rather than any real astrophysical differences between the samples. We calculate the average abundance error of our twelve considered elements for the sample of high fidelity (SNR $>$ 200) field stars, and find that [Na/Fe] exhibits an average error of 0.060 dex, the second-to-largest among these elements after [V/Fe] (0.073 dex). For comparison, the typical average error of these twelve abundances is $\sim$0.03 dex. We conclude that Na is generally poorly measured by the ASPCAP pipeline. 

We also examine the heliocentric velocity $V_{\textrm{helio}}$ as \citet{ness2022} found abundance residual trends with $V_{\textrm{helio}}$ that indicate contamination from ISM features. Our [Na/H] residuals do not display such trends, so we conclude that the intrinsic dispersion differences between KOIs and doppelg\"angers are not due to ISM contamination.

%\mkn{can we check these 5 stars have Na$\_Fe\_$ELEM flag = 0}  

%\textbf{(Should I discuss the heliocentric velocity stuff?)} \mkn{I think worth saying that Ness 2022 finds trends with velocity thought to be due to ISM contamination and so verified that when did selection on the velocity the Na result did not change which indicates that the mesaurement issue is not ISM in the data.}

%These Na outliers motivated the construction of our doppelg\"anger sample also selected on $A_{k}$; there are hints that Na abundances correlate (noisily) with 

\subsection{Higher Quality Sample} 
\begin{figure*} 
\centering
    \vspace*{0.02in}
    \includegraphics[width=0.99\textwidth]{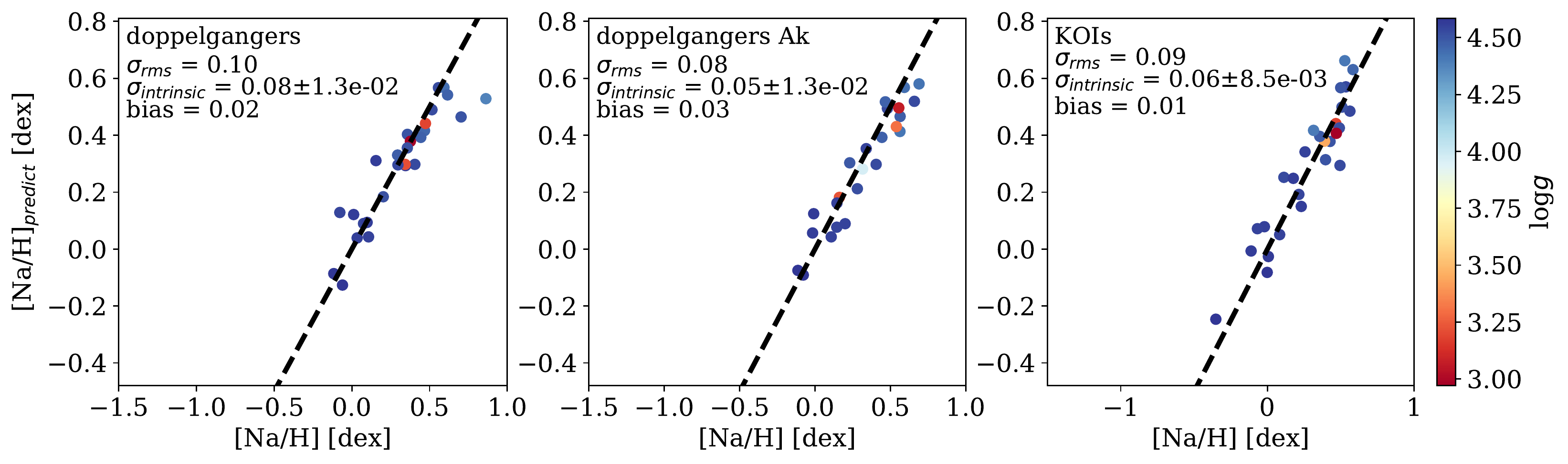}
    %\vspace*{-0.05in}
    \caption{Local linear model-predicted vs. ASPCAP abundances of the higher quality ([X/H]$_{err}$ $<$ 0.07 dex) doppelg\"anger (left), doppelg\"anger $A_{k}$ (middle), and KOI (right) samples for [Na/H]. The points are colored by \logg. The rms difference between the ASPCAP and predicted abundances, intrinsic dispersion, and bias measurements are provided in the upper left corners of each panel. A dashed 1-to-1 line is plotted in all panels for comparison.}
\label{fig:figure7}
\end{figure*}

We examine if the KOI [Na/H] intrinsic dispersion remains much larger than those of the doppelg\"anger samples with more stringent cuts on [X/Fe]$_{\textrm{error}}$. To investigate this, we construct a higher quality sample with cuts on [X/Fe]$_{\textrm{error}}$ $<$ 0.07 dex rather than 0.1 dex. This results in a substantial sample size decrease, from 130 to 27 KOIs. The five outlier stars in predicted and ASPCAP [Na/H] abundance space are removed (Figure \ref{fig:figure7}). The new [Na/H] intrinsic dispersion measurement for the KOIs is $\sigma_{\textrm{intrinsic}}$ $\approx$ 0.062 dex, compared to those of the doppelganger and $A_{k}$ doppelganger samples, $\sigma_{\textrm{intrinsic}}$ $\approx$ 0.085 dex and 0.054 dex, respectively. Median intrinsic dispersions across all abundances are $\sim$0.038 dex and $\sim$0.041 dex for the KOI and doppelg\"anger samples, respectively. This constitutes further evidence that there is likely no systematic difference between the KOI and doppelg\"anger samples, and the initial differences in [Na/H] intrinsic dispersion were driven by targets with large abundance uncertainties.

\subsection{Condensation Temperature Trends} 
\begin{figure*}[t]
    \centering
    \begin{minipage}{0.48\textwidth}
        \centering
        \includegraphics[width=0.99\textwidth]{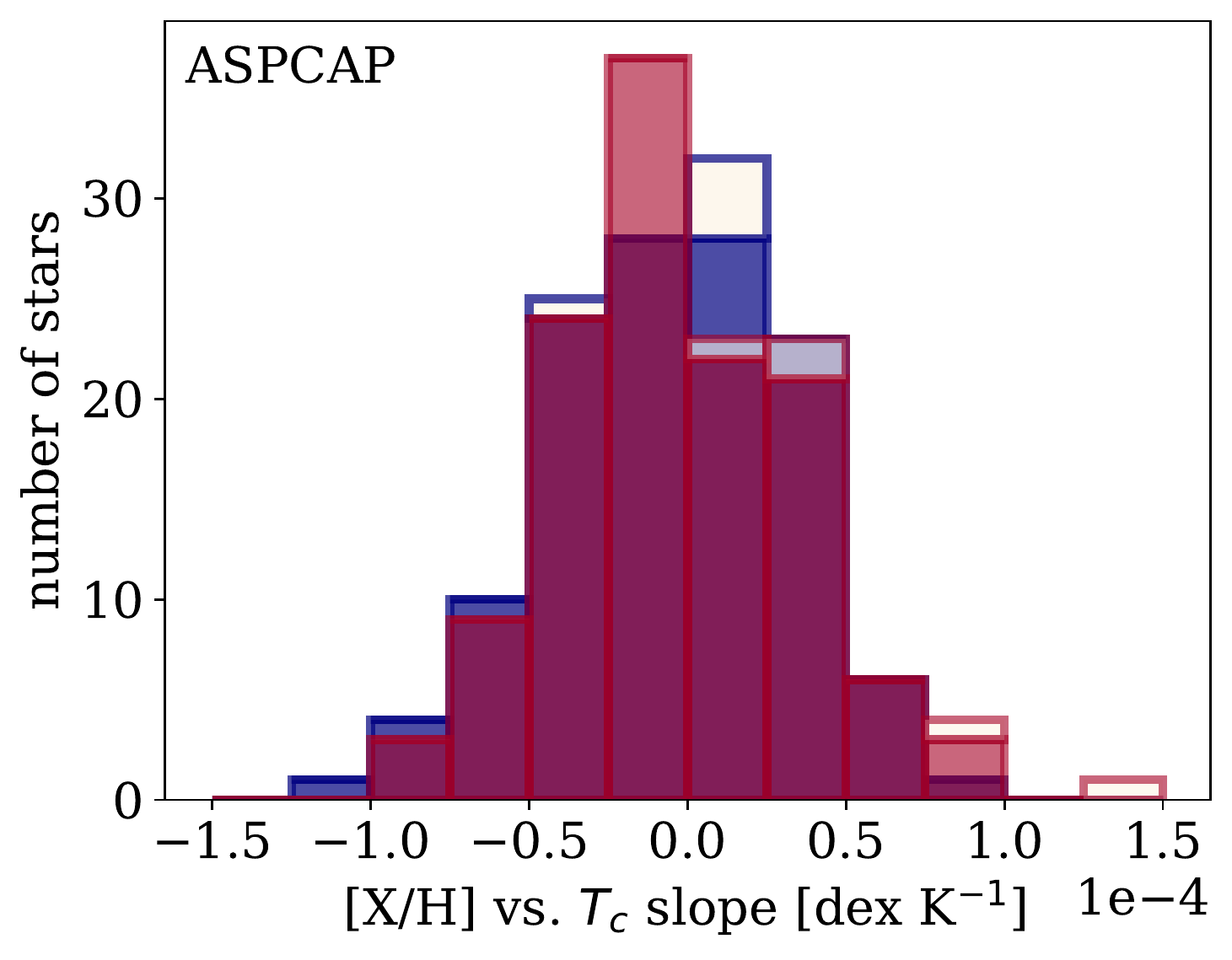} % first figure itself
    \end{minipage}\hfill
    \begin{minipage}{0.49\textwidth}
    \vspace{1mm}
        \centering
        \includegraphics[width=0.99\textwidth]{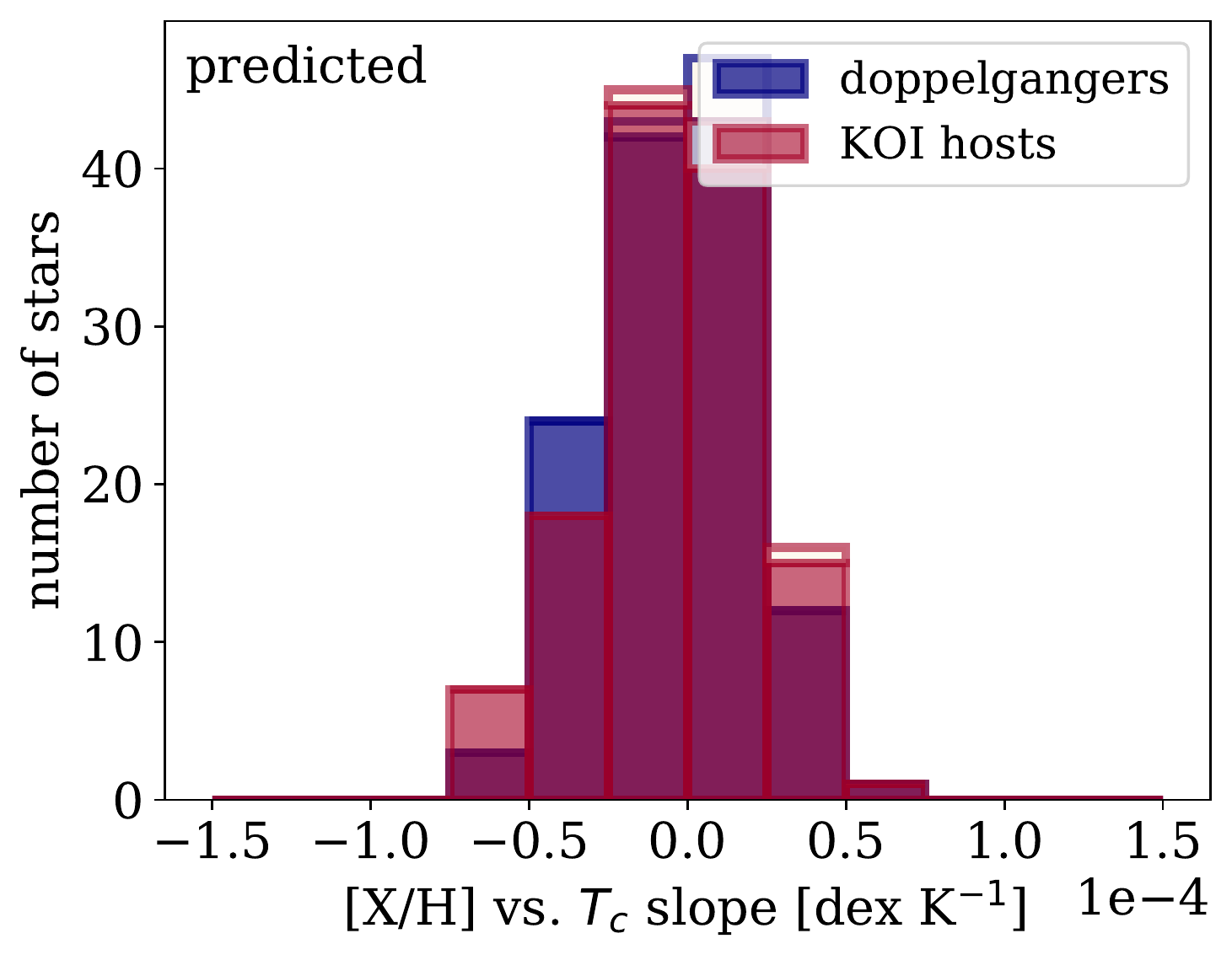} % second figure itself
    \end{minipage}
    \caption{Linear trend slope distributions for the $T_{c}$-ordered abundances of individual stars in the KOI (red) and doppelg\"anger (blue) samples. The distributions for the predicted and ASPCAP abundances are provided in the right and left panels, respectively. The slope contributions from our five Na outlier stars are represented by the transparent regions of the distribution.}
\label{fig:figure8}
\end{figure*}

We further explore possible $T_{c}$ patterns by fitting linear trends to the $T_{c}$-ordered abundances of each KOI and doppelg\"anger star, respectively. We carry this out for the model-predicted and ASPCAP-provided abundances. The distributions of the linear trend slopes are shown in Figure \ref{fig:figure8}. Both distributions are centered on approximately zero, which indicates that there is no excess of $T_{c}$-dependent enrichment or depletion trends.
%\mkn{why is this the case that you don't expect this to be non-zero for the KOIs if this is linked to planet formation?}. 
The predicted and ASPCAP abundance distributions are both approximately symmetric for the KOI and doppelg\"anger samples, but the KOI ASPCAP distribution appears to exhibit a slight excess ($\sim$25 systems) of slopes between approximately $-$0.5$\times$10$^{-5}$ and $-$2.5$\times$10$^{-5}$. We test whether this excess is astrophysical or random in nature by drawing samples from the KOI ASPCAP slope distribution to see how often the excess is reproduced. From 1000 trials, the excess occurs $\sim$10\% of the time, indicating that it is likely a random phenomenon. 

\begin{figure*}[t]
    \centering
    \begin{minipage}{0.48\textwidth}
        \centering
        \includegraphics[width=0.99\textwidth]{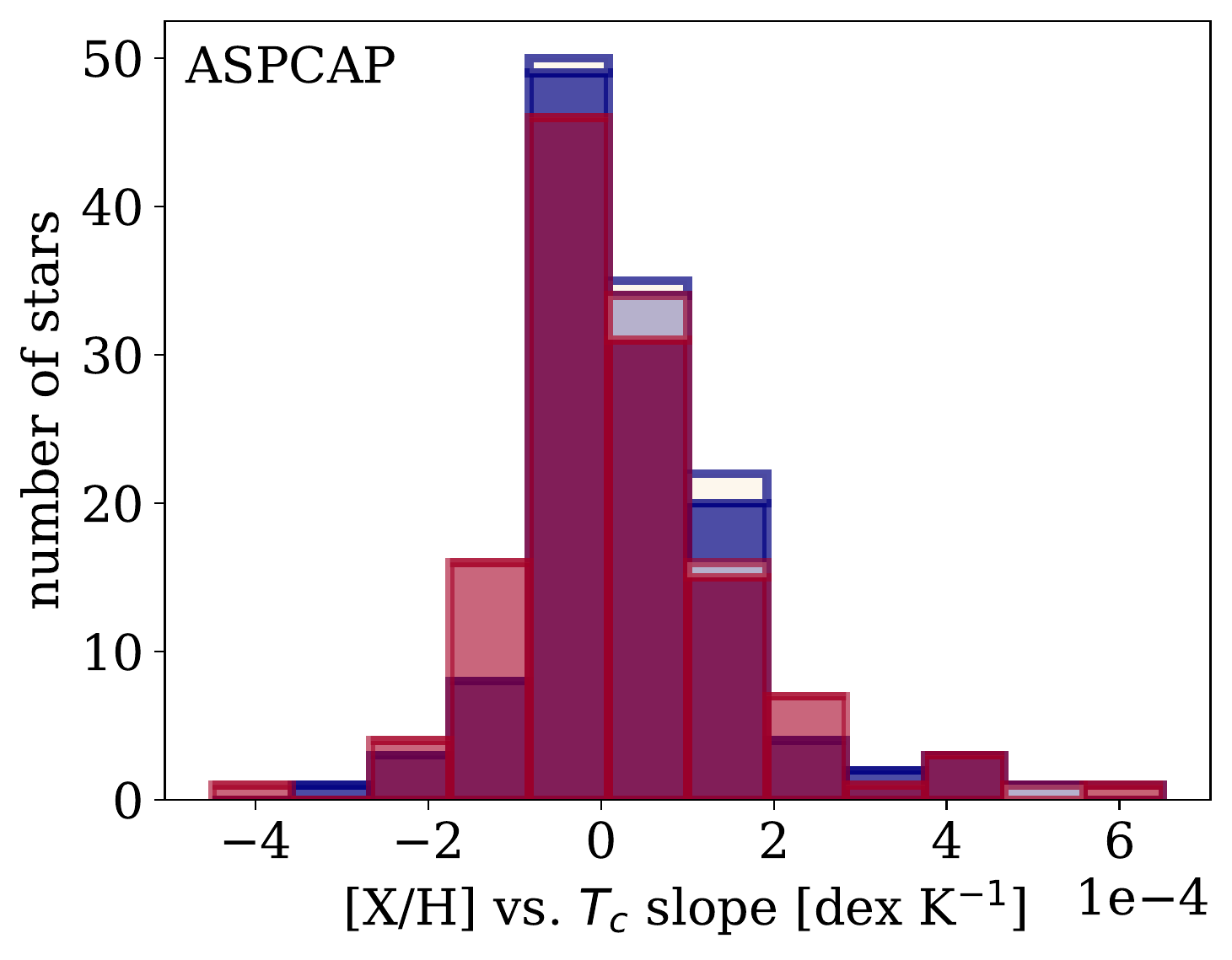} % first figure itself
    \end{minipage}\hfill
    \begin{minipage}{0.48\textwidth}
        \centering
        \includegraphics[width=0.99\textwidth]{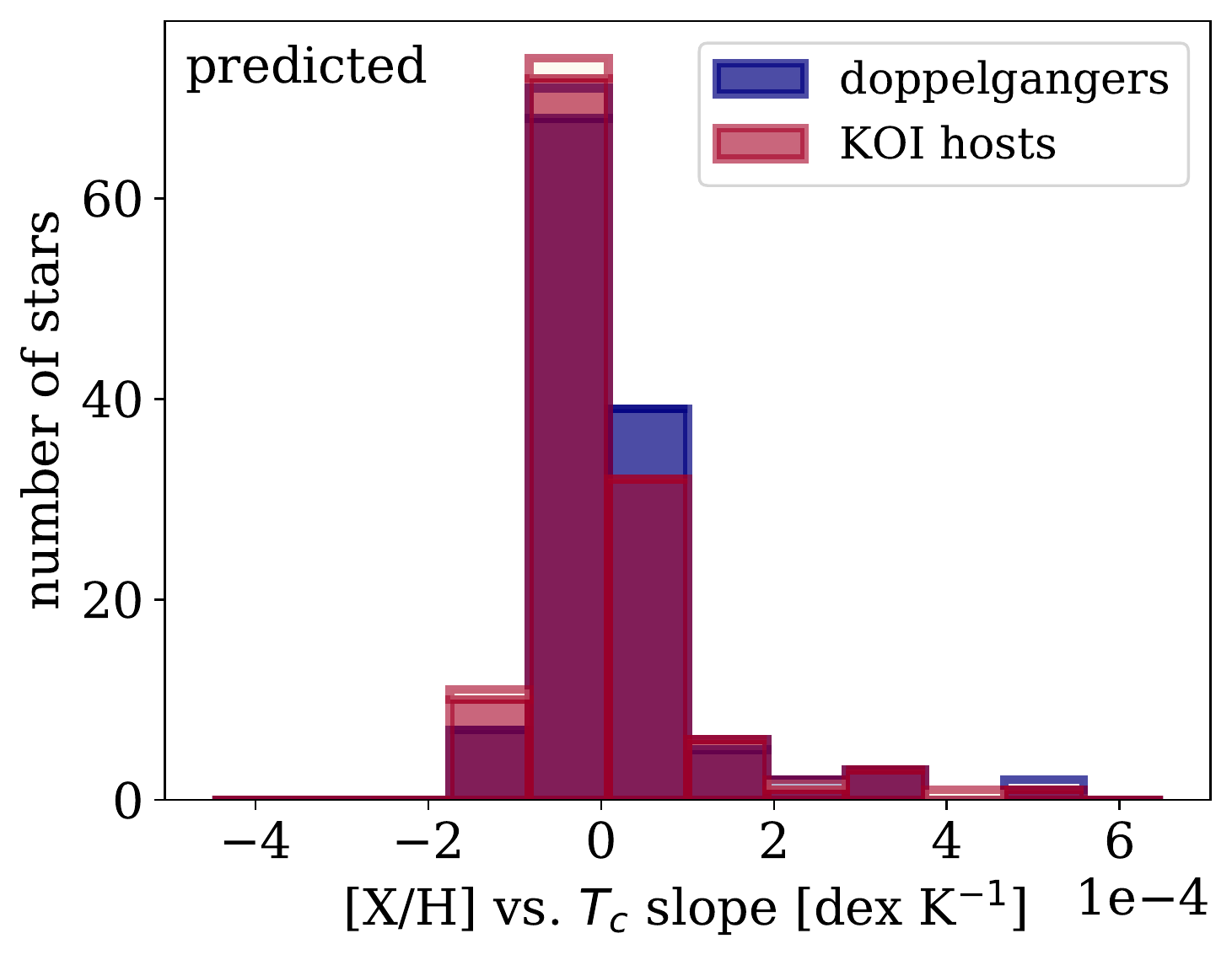} % second figure itself
    \end{minipage}
    \caption{Linear trend slope distributions for the $T_{c}$-ordered abundances of individual stars in the KOI (red) and doppelg\"anger (blue) samples, for only refractory elements ($T_{c}$ $>$ 1000 K). The distributions for the predicted and ASPCAP abundances are provided in the right and left panels, respectively. The slope contributions from our five Na outlier stars are represented by the transparent regions of the distribution.}
\label{fig:figure9}
\end{figure*}

We construct similar distributions of linear trend slopes resulting from fits to $T_{c}$-ordered abundances, but only considering elements with $T_{c}$ $>$ 1000 K (Mn, Cr, Si, Ni, V, Ca, Ti, Al). Elements are refractory rather than volatile above this temperature, and populate the steepest regions of abundance versus $T_{c}$ trends in patterns exhibiting refractory enhancement or depletion. \citep{melendez2009,ramirez2009,bedell2018}. This is similar to the analysis presented in \citet{nibauer2021}, which assessed $T_{c}$ trends of elements with $T_{c}$ $>$ 900 K. The resulting slope distributions for the predicted and ASPCAP abundances are shown in Figure \ref{fig:figure9}. Both the ASPCAP and predicted abundance distributions for the KOI and doppelg\"anger samples exhibit a tail of slopes towards the right of the distribution center that appears to peak at $\sim$4$\times$10$^{-4}$ dex K$^{-1}$. A similar secondary peak was found by \citet{nibauer2021} in their $T_{c}$ slope distribution for $>$900 K elements from APOGEE DR16 data. This indicates that our data reproduces the $T_{c}$ patterns in field stars, which is reassuring. However, we find fewer stars in the secondary peak at positive gradients compared to \citet{nibauer2021}. This is potentially explained by our different stellar samples that span different evolutionary states; \citet{nibauer2021} examined stars across a narrow range of the main sequence whereas we study stars across the main sequence and red giant branch. 

Another interesting feature in our ASPCAP abundance distributions is a small tail towards negative slopes that stretches beyond $-$2$\times$10$^{-4}$ dex K$^{-1}$ (Figure \ref{fig:figure9}, left panel). This tail of negative slope values is not apparent in our predicted abundance distributions (Figure \ref{fig:figure9}, right panel), or the \citet{nibauer2021} results. We calculate that $\sim$10\% and $\sim$7.7\% of the KOI and doppelg\"anger ASPCAP distributions are below $-$2$\times$10$^{-4}$ dex K$^{-1}$, respectively, and therefore compose the negative slope tail. The presence of this tail in the ASPCAP abundance distribution but not the predicted abundance distribution may indicate that there is abundance information not fully captured by (Fe, Mg) alone that may alter dex vs. $T_{c}$ trends at the most negative slope regions. The absence of this tail in the \citet{nibauer2021} abundances, which are comparable to ASPCAP abundances, may again be due to evolutionary state differences in our respective stellar samples.

%This secondary peak is not visible in the $T_{c}$ slope distribution of ASPCAP abundances for KOIs or doppelg\"angers. The \citet{nibauer2021} abundance data is generated from a hierarchical mixture model that assumes the abundance values are linear functions of $T_{c}$. The fact that our predicted, but not ASPCAP, abundance distributions reproduce the secondary peak visible in the \citet{nibauer2021} data could indicate that abundance information not fully captured by (Fe, Mg) or $T_{c}$ may alter $T_{c}$ trends. 

%Still, the secondary peak in our predicted abundance distribution is slight; we conclude that there is no striking difference between the ASPCAP and predicted abundance $T_{c}$ $>$ 1000 K distributions.

\begin{figure*}[t]
    \centering
    \begin{minipage}{0.95\textwidth}
        \centering
        \includegraphics[width=0.99\textwidth]{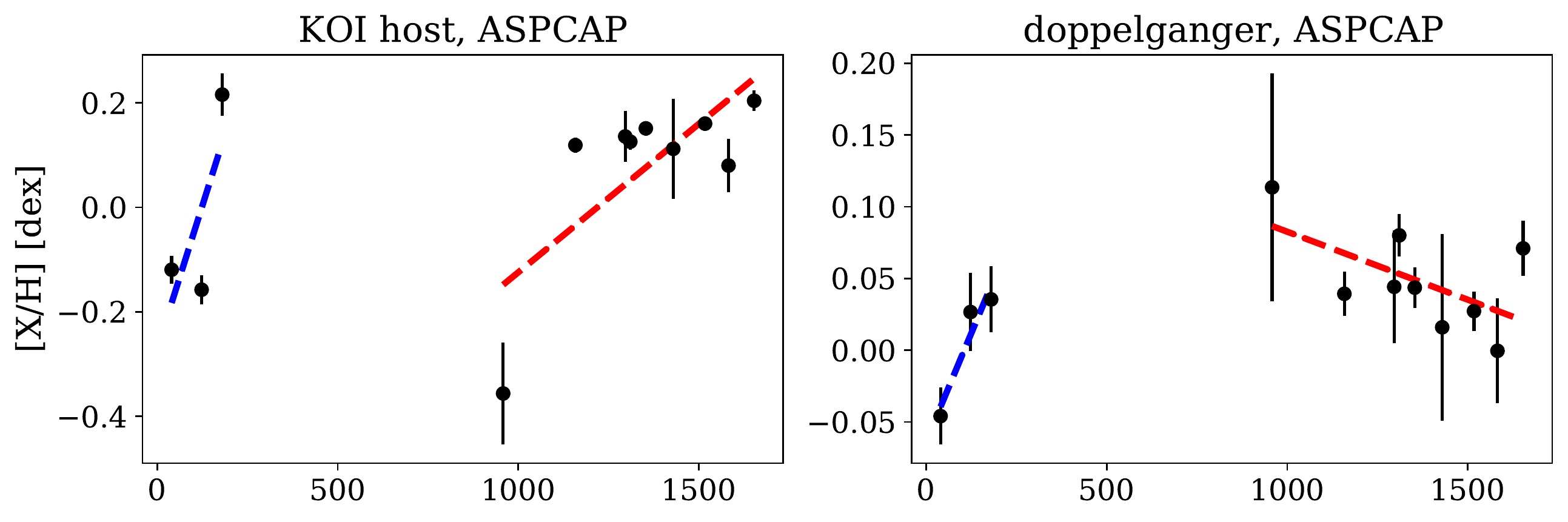} % first figure itself
    \end{minipage}\hfill
    \hspace{10mm}
    \begin{minipage}{0.94\textwidth}
        \centering
        \hspace*{0.1in}
        \includegraphics[width=0.99\textwidth]{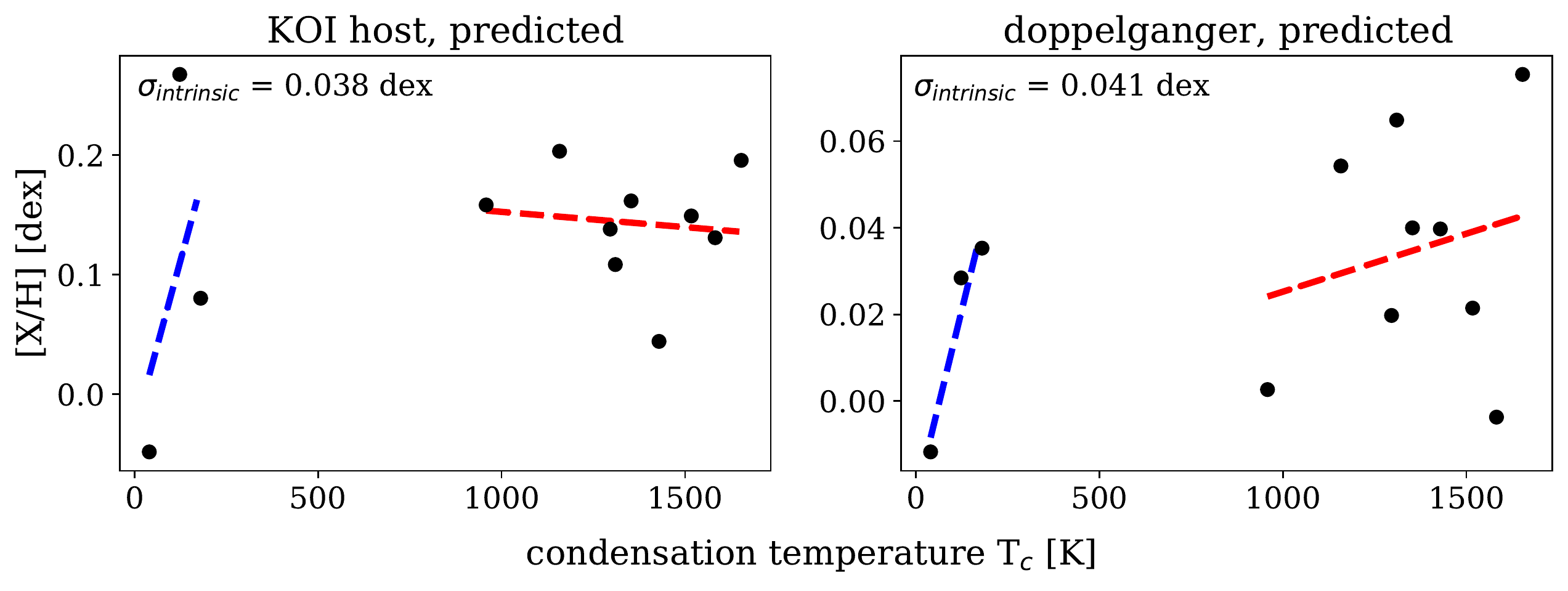} % second figure itself
    \end{minipage}
    \caption{Examples of separate linear trends fitted to volatile (blue) and refractory (red) abundances ordered by $T_{c}$ for our KOIs and corresponding doppelg\"angers. The boundary between volatile and refractory elements is set to $T_{c}$ = 1000 K. Linear fits to the ASPCAP and predicted abundances are provided in the upper and lower panels, respectively. There are no individual errors associated with the predicted abundances, but the median intrinsic dispersion values for the KOI and doppelg\"anger samples across all abundances are shown in the upper left corners of the predicted abundance panels.}
\label{fig:figure10}
\end{figure*}

\begin{figure*}[t]
    \centering
    \begin{minipage}{0.48\textwidth}
        \centering
        \includegraphics[width=0.99\textwidth]{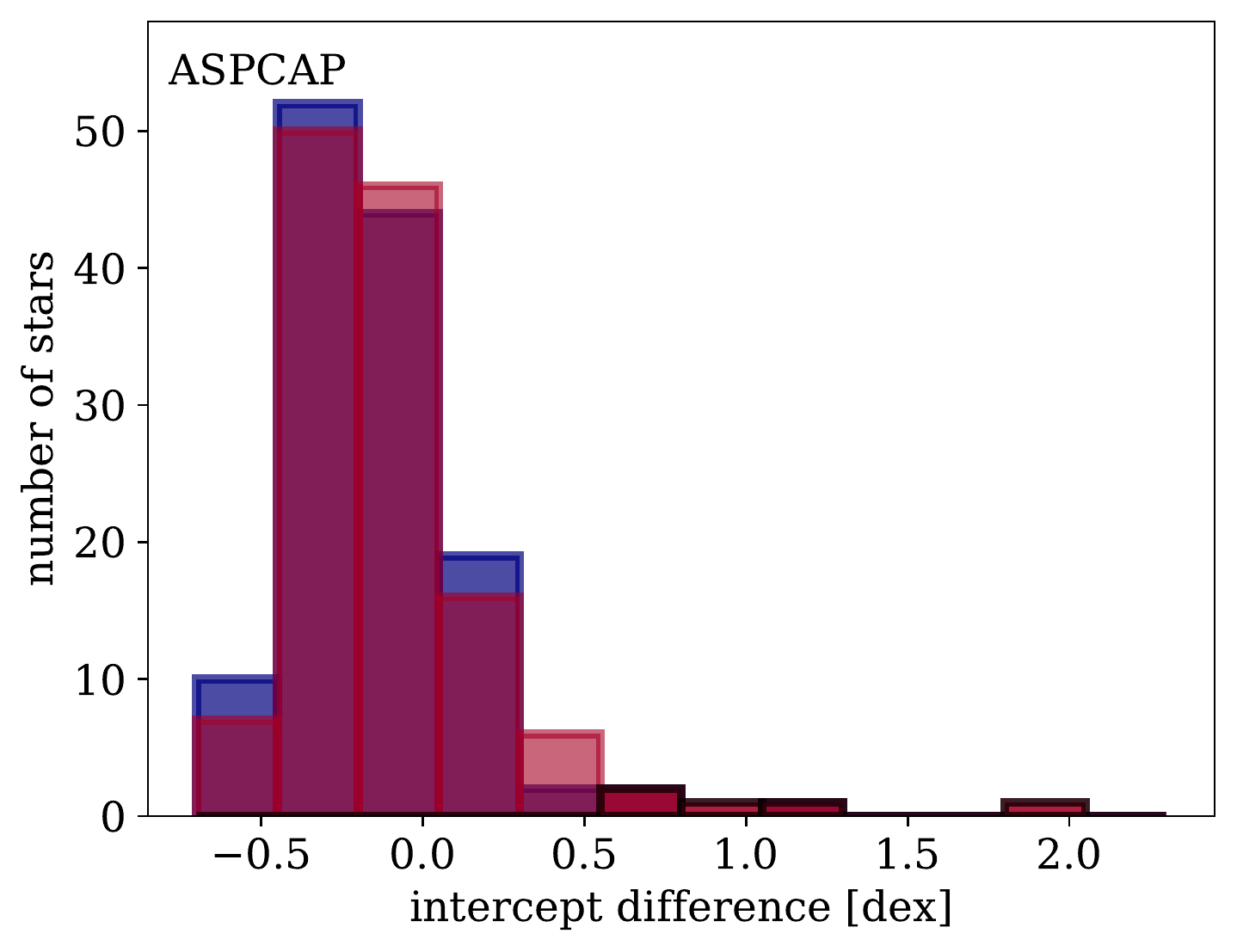} % first figure itself
    \end{minipage}\hfill
    \begin{minipage}{0.48\textwidth}
        \centering
        \includegraphics[width=0.99\textwidth]{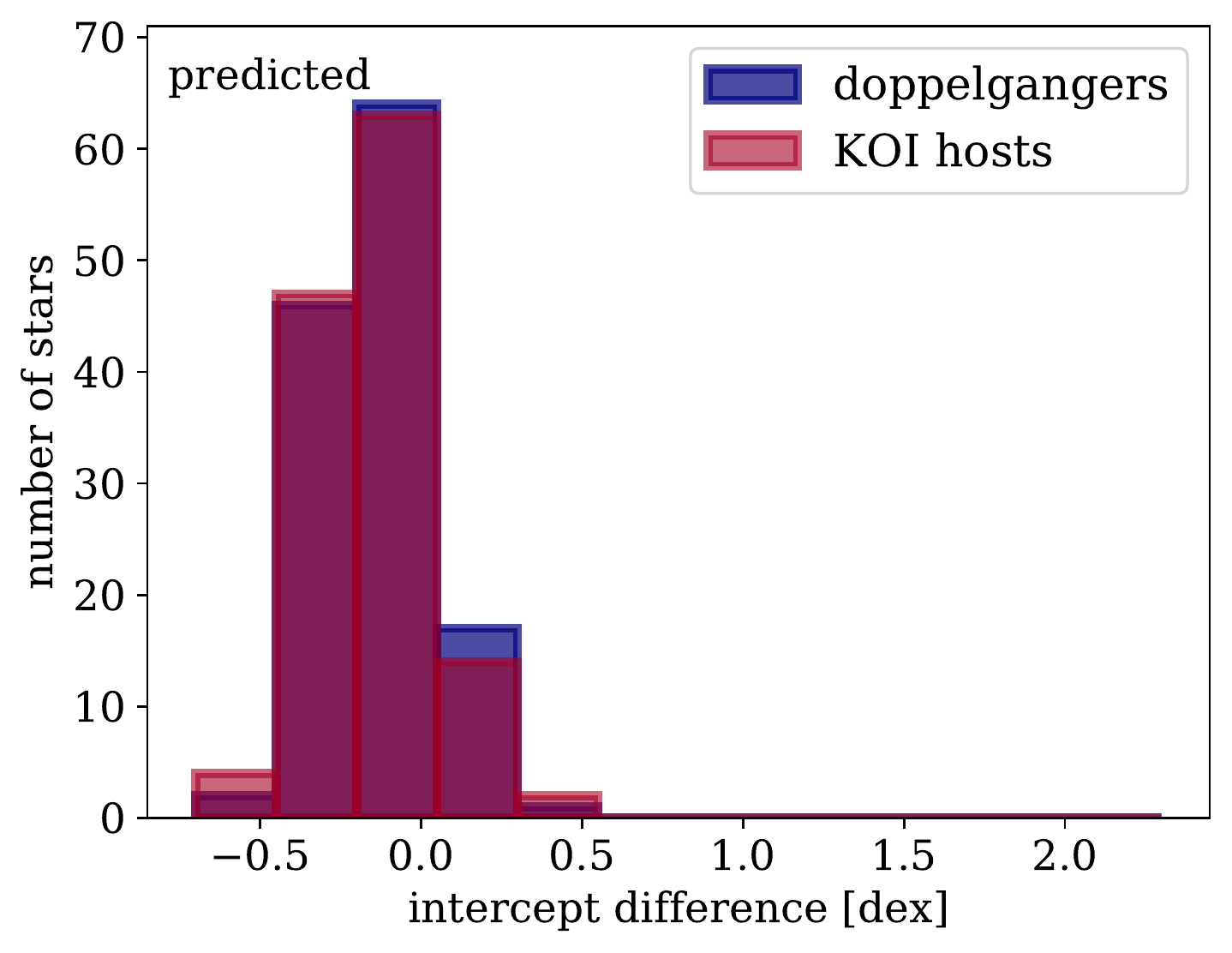} % second figure itself
    \end{minipage}
    \caption{Intercept difference (refractory$-$volatile) distributions from linear trend fits to the $T_{c}$-ordered refractory and volatile abundances of individual stars in the KOI (red) and doppelg\"anger (blue) samples. The distributions for the predicted and ASPCAP abundances are provided in the right and left panels, respectively. The contributions corresponding to the KOIs with the top five largest intercept differences are outlined in black in the KOI ASPCAP distribution. Four out of these five KOIs overlap with the five KOIs that are outliers in predicted and ASPCAP [Na/H] space.}
\label{fig:figure11}
\end{figure*}

$T_{c}$ patterns can also be examined by splitting abundances into volatile and refractory groups, and fitting individual linear trends to both sets. This was done by \citet{bedell2018} for a sample of solar twins (see their Figure 4). Stars with enrichment trends will exhibit steeper linear fits to abundances with $T_{c}$ $>$ 1000 K compared to abundances with $T_{c}$ $<$ 1000 K, while the opposite will be true for depletion trends. We carry out this analysis for our KOIs and their doppelg\"angers, and provide examples of our linear trend fits in Figure \ref{fig:figure10}. Because strong enrichment results in steeper refractory trends, the linear fits will have lower intercept values. Thus, enrichment pattern strength can be likened to the difference in volatile and refractory linear fit intercepts. We plot the distributions of these intercept differences in Figure \ref{fig:figure11}. The distribution corresponding to ASPCAP abundances exhibits a tail towards higher intercept differences that is not present in the distribution derived from predicted abundances. We examine the ASPCAP $T_{c}$ trends for the KOIs with the top five largest intercept differences, and find that they have anomalously low measured [Na/H] (ranging from $-$0.23 dex to $-$1.29 dex) that are $\gtrsim$0.2 dex below the other measured abundances. The associated errors on measured [Na/H] are large (0.074$-$0.94 dex). In addition, four out of the five KOIs with largest intercept differences overlap with the five KOIs that are outliers in predicted and ASPCAP [Na/H] space (Figure \ref{fig:figure4}, [Na/H] KOI panel). This is further evidence that the [Na/H] intrinsic dispersion differences in the initial sample selected on [X/Fe]$_{\textrm{error}}$ $<$ 0.1 dex are the result of large abundance uncertainties. We conclude that if there are underlying differences in the individual abundance $T_{c}$ trends for the KOI and doppelg\"anger samples at fixed evolutionary state, [Fe/H], and [Mg/H], they are marginal. To be detected, these differences must exceed the sensitivity of our predicted abundances, which is typically $\sigma_{\textrm{intrinsic}}$ $\approx$ 0.038 dex and 0.041 dex for the KOIs and doppelg\"angers, respectively.

%\mkn{We conclude that if there is any underlying difference in the populations of KOIs and their doppelgangers in individual abundances with condensation temperature trends at fixed evolutionary state and [Fe/H], [Mg/H], it is exceptionally marginal. This must exceed the sensitivity of our measurements (which is typically $<$ 0.02 (?put in right number here?) dex on average for the elements - number is median abundance uncertainty across all elements for sample} 

%\mkn{suggest to remove this below, replaced as above - I think this statement below isn't quite correct as we can still and have carried out very reliable and robust Tc analysis here - we just don't see any differences. Because we are pooling stars the precision of the distribution is also very high so a pretty solid test that there are no real differences}
%: Stellar samples with higher abundance precisions are thus needed for reliable $T_{c}$ trend analyses. 

\section{Discussion} \label{sec:discussion}

The planet-metallicity correlation remains the only proven connection between host star chemistry and planet properties. %\textbf{(Could mention trends with abundances beyond [Fe/H], but they are weak and hard to disentangle from galactic chemical evolution e.g., Wilson+2022)}
We demonstrate that after removing the effects of evolutionary state and metallicity from two primary sources (Fe, Mg), the individual abundances of confirmed/candidate planet hosts (KOIs) and a reference doppelg\"anger set with unknown planet membership are indistinguishable. More specifically, we compute model-measurement abundance residuals from ASPCAP and predicted abundances using a four-parameter model (\teff, \logg, [Fe/H], [Mg/H]), and find that there are no differences in residual structure between the KOI and doppelg\"anger samples. We calculate the median intrinsic dispersion across all analyzed elements other than (Fe, Mg) to be $\sigma_{\textrm{intrinsic}}$ $\approx$ 0.038 dex and 0.041 dex for the KOI and doppelg\"anger samples, respectively, which can be taken as the minimum abundance precision required for discerning individual abundance signatures related to planet formation. 

Because we do not know the planet membership of our doppelg\"anger sample, some doppelg\"anger stars may be planet hosts. This is plausible because large planet discovery surveys such as the Kepler and the Transiting Exoplanet Survey Satellite (TESS) missions have revealed that planets are common. Using Kepler DR25, \citet{hsu2019} recently calculated an upper limit occurrence rate of 0.27 planets per star for 0.5$-$16 $R_{\oplus}$ planets around FGK dwarfs. Breaking occurrence rates by planet architectures reveals that the majority of these planets are small ($R$ = 1$-$4 $R_{\oplus}$) and generally classified as super-Earths and sub-Neptunes (e.g., \citealt{burke2015,zhu2018,bryson2021}). If a significant fraction of our doppelg\"anger set consists of planet hosts, it makes sense that the abundance distributions of our KOI and doppelg\"anger samples are indistinguishable at fixed (Fe, Mg) and evolutionary state.

To reliably examine abundance differences between planet hosts and reference doppelg\"anger stars drawn from the field, none of the reference stars should host planets. Unfortunately, constructing a sample of doppelg\"anger stars that we know lack planets is difficult. This would require extensive monitoring of targets with Doppler planet search surveys to ensure that there are no radial velocity signals indicative of planets. Carrying out such observations for an entire reference set of stars would be time and resource intensive. However, certain planet populations can be ruled out with minimal telescope time; close-in giant planets are more easily detected in radial velocity and transit data without long cadence compared to smaller planets on longer orbits. In addition, close-in giants are intrinsically rare. Radial velocity surveys produce hot Jupiter ($P$ $<$ 10 days) occurrence rates of $\sim$0.8$-$1.2\% around solar-like stars (e.g., \citealt{mayor2011,wright2012,wittenmyer2020}), and transit surveys yield even smaller occurrence rates of $\sim$0.4$-$0.6\% \citep{howard2012,fressin2013,petigura2018,kunimoto2020}. These rates are still small for warm Jupiters ($P$ $<$ 50 days), with estimates of $\sim$1.3\%. They remain small for hot and warm sub-Saturns ($R$ = 4$-$8 $R_{\oplus}$) as well, which have occurrence rate estimates of $\sim$0.4\% and $\sim$2.3\%, respectively \citep{howard2012}. Thus, constructing a reference sample without close-in giant hosts is feasible. We hope to examine close-in giants in future studies, but this will require another planet host sample as only 18 of our KOIs host confirmed/candidate hot/warm sub-Saturn to Jupiter-sized planets according to the standard definition ($R$ $>$ 4 $R_{\oplus}$ and $P$ $<$ 100 days).

Previous studies have found interesting abundance differences between stars that host and do not host close-in giants. For example, \citet{melendez2009} determined that the Sun exhibits a refractory depletion trend with $T_c$ relative to eleven solar twins from the \emph{Hipparcos} catalog, as well as four solar analogs with close-in giant planets. However, six other solar analogs lacking close-in giants as verified by radial velocity monitoring show the solar depletion trend 50$-$70\% of the time. One potential explanation for the solar pattern is sequestration of rocky material in the terrestrial planets, and late (10$-$25 Myr) accretion of dust-depleted gas once the solar convective zone began shrinking to its current mass fraction ($\sim$2\%, \citealt{hughes2007}). Another explanation is that all solar twins and most solar analogs lacking close-in giants engulfed planetary material at late times ($>$25 Myr), once their convective zones were thin. This scenario would produce refractory enrichment in stellar photospheres. However, it assumes that most solar-like stars are depleted in refractories (at least in the absence of events like planet engulfment), and more recent abundance studies of larger Sun-like samples show that this is not the case (e.g., \citealt{bedell2018}). Either way, the findings of \citet{melendez2009} suggest that close-in giant planets play a role in altering host star abundances. While their results defy a clear explanation, a larger sample of close-in giant hosts and reference stars lacking close-in giants could be leveraged to examine these trends more closely.

The KOI and doppelg\"anger median abundance prediction intrinsic dispersions are $\sim$0.038 dex and $\sim$0.041 dex, respectively. These values can be considered the upper limit of abundance precision needed to discern planet formation signatures in the elemental abundance patterns of host stars. Planet formation processes can exceed these levels in rare cases, such as the reported planet engulfment detection in the HD 240429-30 system ($\sim$0.2 dex, \citealt{oh2018}). Planet hosts may also be born with different abundances compared to stars without planets. The planet-metallicity correlation indicates that this is true for at least [Fe/H]. Such primordial abundance deviations must also exceed our intrinsic dispersion levels to be detectable.

Our KOI and doppelg\"anger residual abundance distributions are indistinguishable, which yields two possibilities: (1) our reference doppelg\"anger set includes too many planet hosts, or (2) primordial or post-birth abundance patterns related to planet formation in our samples are below detectable levels. We can tackle the first possibility by focusing on more easily detectable planet architectures, namely close-in giants as discussed earlier. The second possibility could be addressed with higher-precision abundances from advances in spectral synthesis pipelines and/or line lists (e.g., \citealt{schuler2011,liu2018,bedell2014}), or from spectrographs with higher resolving power (e.g., \citealt{adibekyan2020}). Many stars in our KOI and doppelg\"anger samples have abundance uncertainties that exceed our intrinsic dispersion values. Large uncertainties are the root cause of the particularly poorly measured Na abundances for the five outlier stars in our initial sample selected on [X/H]$_{\textrm{err}}$ $<$ 0.1 dex. Upgrades to the ASPCAP pipeline, such as improved line lists and advances to the spectral synthesis pipeline, may improve APOGEE abundance precisions in the years to come. %\mkn{I think the discussion is excellent}

\section*{Acknowledgements}
\begin*{}

A.B. acknowledges funding from the National Science Foundation Graduate Research Fellowship under Grant No. DGE1745301. This work benefited from involvement in ExoExplorers, which is sponsored by the Exoplanets Program Analysis Group (ExoPAG) and NASA’s Exoplanet Exploration Program Office (ExEP). E.C.C. gratefully acknowledges support for this work provided by NASA through the NASA Hubble Fellowship Program grant No. HST-HF2-51502.001-A awarded by the Space Telescope Science Institute, which is operated by the Association of Universities for Research in Astronomy, Inc., for NASA, under contract NAS5-26555.
\software{\texttt{numpy} \citep{numpy}, \texttt{matplotlib} \citep{matplotlib}, \texttt{pandas} \citep{pandas}, \texttt{scipy} \citep{scipy}, \texttt{scikit-learn} \citep{scikit-learn}, \texttt{astropy} \citep{astropy:2013, astropy:2018}}
\end*{}

\bibliography{mybib}{}

\begin{thebibliography}{}
\expandafter\ifx\csname natexlab\endcsname\relax\def\natexlab#1{#1}\fi
\providecommand{\url}[1]{\href{#1}{#1}}
\providecommand{\dodoi}[1]{doi:~\href{http://doi.org/#1}{\nolinkurl{#1}}}
\providecommand{\doeprint}[1]{\href{http://ascl.net/#1}{\nolinkurl{http://ascl.net/#1}}}
\providecommand{\doarXiv}[1]{\href{https://arxiv.org/abs/#1}{\nolinkurl{https://arxiv.org/abs/#1}}}

\bibitem[{{Abdurro'uf} {et~al.}(2022){Abdurro'uf}, {Accetta}, {Aerts}, {Silva
  Aguirre}, {Ahumada}, {Ajgaonkar}, {Filiz Ak}, {Alam}, {Allende Prieto},
  {Almeida}, {Anders}, {Anderson}, {Andrews}, {Anguiano}, {Aquino-Ort{\'\i}z},
  {Arag{\'o}n-Salamanca}, {Argudo-Fern{\'a}ndez}, {Ata}, {Aubert},
  {Avila-Reese}, {Badenes}, {Barb{\'a}}, {Barger}, {Barrera-Ballesteros},
  {Beaton}, {Beers}, {Belfiore}, {Bender}, {Bernardi}, {Bershady}, {Beutler},
  {Bidin}, {Bird}, {Bizyaev}, {Blanc}, {Blanton}, {Boardman}, {Bolton},
  {Boquien}, {Borissova}, {Bovy}, {Brandt}, {Brown}, {Brownstein}, {Brusa},
  {Buchner}, {Bundy}, {Burchett}, {Bureau}, {Burgasser}, {Cabang}, {Campbell},
  {Cappellari}, {Carlberg}, {Wanderley}, {Carrera}, {Cash}, {Chen}, {Chen},
  {Cherinka}, {Chiappini}, {Choi}, {Chojnowski}, {Chung}, {Clerc}, {Cohen},
  {Comerford}, {Comparat}, {da Costa}, {Covey}, {Crane}, {Cruz-Gonzalez},
  {Culhane}, {Cunha}, {Dai}, {Damke}, {Darling}, {Davidson}, {Davies},
  {Dawson}, {De Lee}, {Diamond-Stanic}, {Cano-D{\'\i}az}, {S{\'a}nchez},
  {Donor}, {Duckworth}, {Dwelly}, {Eisenstein}, {Elsworth}, {Emsellem},
  {Eracleous}, {Escoffier}, {Fan}, {Farr}, {Feng}, {Fern{\'a}ndez-Trincado},
  {Feuillet}, {Filipp}, {Fillingham}, {Frinchaboy}, {Fromenteau}, {Galbany},
  {Garc{\'\i}a}, {Garc{\'\i}a-Hern{\'a}ndez}, {Ge}, {Geisler}, {Gelfand},
  {G{\'e}ron}, {Gibson}, {Goddy}, {Godoy-Rivera}, {Grabowski}, {Green},
  {Greener}, {Grier}, {Griffith}, {Guo}, {Guy}, {Hadjara}, {Harding},
  {Hasselquist}, {Hayes}, {Hearty}, {Hern{\'a}ndez}, {Hill}, {Hogg},
  {Holtzman}, {Horta}, {Hsieh}, {Hsu}, {Hsu}, {Huber}, {Huertas-Company},
  {Hutchinson}, {Hwang}, {Ibarra-Medel}, {Chitham}, {Ilha}, {Imig}, {Jaekle},
  {Jayasinghe}, {Ji}, {Johnson}, {Jones}, {J{\"o}nsson}, {Katkov}, {Khalatyan},
  {Kinemuchi}, {Kisku}, {Knapen}, {Kneib}, {Kollmeier}, {Kong}, {Kounkel},
  {Kreckel}, {Krishnarao}, {Lacerna}, {Lane}, {Langgin}, {Lavender}, {Law},
  {Lazarz}, {Leung}, {Leung}, {Lewis}, {Li}, {Li}, {Lian}, {Liang}, {Lin},
  {Lin}, {Lin}, {Lintott}, {Long}, {Longa-Pe{\~n}a}, {L{\'o}pez-Cob{\'a}},
  {Lu}, {Lundgren}, {Luo}, {Mackereth}, {de la Macorra}, {Mahadevan},
  {Majewski}, {Manchado}, {Mandeville}, {Maraston}, {Margalef-Bentabol},
  {Masseron}, {Masters}, {Mathur}, {McDermid}, {Mckay}, {Merloni},
  {Merrifield}, {Meszaros}, {Miglio}, {Di Mille}, {Minniti}, {Minsley},
  {Monachesi}, {Moon}, {Mosser}, {Mulchaey}, {Muna}, {Mu{\~n}oz}, {Myers},
  {Myers}, {Nadathur}, {Nair}, {Nandra}, {Neumann}, {Newman}, {Nidever},
  {Nikakhtar}, {Nitschelm}, {O'Connell}, {Garma-Oehmichen}, {Luan Souza de
  Oliveira}, {Olney}, {Oravetz}, {Ortigoza-Urdaneta}, {Osorio}, {Otter},
  {Pace}, {Padilla}, {Pan}, {Pan}, {Parikh}, {Parker}, {Peirani}, {Pe{\~n}a
  Ram{\'\i}rez}, {Penny}, {Percival}, {Perez-Fournon}, {Pinsonneault},
  {Poidevin}, {Poovelil}, {Price-Whelan}, {B{\'a}rbara de Andrade Queiroz},
  {Raddick}, {Ray}, {Rembold}, {Riddle}, {Riffel}, {Riffel}, {Rix}, {Robin},
  {Rodr{\'\i}guez-Puebla}, {Roman-Lopes}, {Rom{\'a}n-Z{\'u}{\~n}iga}, {Rose},
  {Ross}, {Rossi}, {Rubin}, {Salvato}, {S{\'a}nchez}, {S{\'a}nchez-Gallego},
  {Sanderson}, {Santana Rojas}, {Sarceno}, {Sarmiento}, {Sayres}, {Sazonova},
  {Schaefer}, {Schiavon}, {Schlegel}, {Schneider}, {Schultheis}, {Schwope},
  {Serenelli}, {Serna}, {Shao}, {Shapiro}, {Sharma}, {Shen}, {Shetrone}, {Shu},
  {Simon}, {Skrutskie}, {Smethurst}, {Smith}, {Sobeck}, {Spoo}, {Sprague},
  {Stark}, {Stassun}, {Steinmetz}, {Stello}, {Stone-Martinez},
  {Storchi-Bergmann}, {Stringfellow}, {Stutz}, {Su}, {Taghizadeh-Popp},
  {Talbot}, {Tayar}, {Telles}, {Teske}, {Thakar}, {Theissen}, {Tkachenko},
  {Thomas}, {Tojeiro}, {Hernandez Toledo}, {Troup}, {Trump}, {Trussler},
  {Turner}, {Tuttle}, {Unda-Sanzana}, {V{\'a}zquez-Mata}, {Valentini},
  {Valenzuela}, {Vargas-Gonz{\'a}lez}, {Vargas-Maga{\~n}a}, {Alfaro},
  {Villanova}, {Vincenzo}, {Wake}, {Warfield}, {Washington}, {Weaver},
  {Weijmans}, {Weinberg}, {Weiss}, {Westfall}, {Wild}, {Wilde}, {Wilson},
  {Wilson}, {Wilson}, {Wolf}, {Wood-Vasey}, {Yan}, {Zamora}, {Zasowski},
  {Zhang}, {Zhao}, {Zheng}, {Zheng}, \& {Zhu}}]{abdurrouf2022}
{Abdurro'uf}, {Accetta}, K., {Aerts}, C., {et~al.} 2022, \apjs, 259, 35,
  \dodoi{10.3847/1538-4365/ac4414}

\bibitem[{{Adibekyan} {et~al.}(2016){Adibekyan}, {Delgado-Mena}, {Figueira},
  {Sousa}, {Santos}, {Faria}, {Gonz{\'a}lez Hern{\'a}ndez}, {Israelian},
  {Harutyunyan}, {Su{\'a}rez-Andr{\'e}s}, \& {Hakobyan}}]{adibekyan2016}
{Adibekyan}, V., {Delgado-Mena}, E., {Figueira}, P., {et~al.} 2016, \aap, 591,
  A34, \dodoi{10.1051/0004-6361/201628453}

\bibitem[{{Adibekyan} {et~al.}(2020){Adibekyan}, {Sousa}, {Santos}, {Figueira},
  {Allende Prieto}, {Delgado Mena}, {Gonz{\'a}lez Hern{\'a}ndez}, {de Laverny},
  {Recio-Blanco}, {Campante}, {Tsantaki}, {Hakobyan}, {Oshagh}, {Faria},
  {Bergemann}, {Israelian}, \& {Boulet}}]{adibekyan2020}
{Adibekyan}, V., {Sousa}, S.~G., {Santos}, N.~C., {et~al.} 2020, \aap, 642,
  A182, \dodoi{10.1051/0004-6361/202038793}

\bibitem[{{Adibekyan} {et~al.}(2012{\natexlab{a}}){Adibekyan}, {Delgado Mena},
  {Sousa}, {Santos}, {Israelian}, {Gonz{\'a}lez Hern{\'a}ndez}, {Mayor}, \&
  {Hakobyan}}]{adibekyan2012b}
{Adibekyan}, V.~Z., {Delgado Mena}, E., {Sousa}, S.~G., {et~al.}
  2012{\natexlab{a}}, \aap, 547, A36, \dodoi{10.1051/0004-6361/201220167}

\bibitem[{{Adibekyan} {et~al.}(2012{\natexlab{b}}){Adibekyan}, {Santos},
  {Sousa}, {Israelian}, {Delgado Mena}, {Gonz{\'a}lez Hern{\'a}ndez}, {Mayor},
  {Lovis}, \& {Udry}}]{adibekyan2012a}
{Adibekyan}, V.~Z., {Santos}, N.~C., {Sousa}, S.~G., {et~al.}
  2012{\natexlab{b}}, \aap, 543, A89, \dodoi{10.1051/0004-6361/201219564}

\bibitem[{{Alibert} {et~al.}(2011){Alibert}, {Mordasini}, \&
  {Benz}}]{alibert2011}
{Alibert}, Y., {Mordasini}, C., \& {Benz}, W. 2011, \aap, 526, A63,
  \dodoi{10.1051/0004-6361/201014760}

\bibitem[{{Astropy Collaboration} {et~al.}(2013){Astropy Collaboration},
  {Robitaille}, {Tollerud}, {Greenfield}, {Droettboom}, {Bray}, {Aldcroft},
  {Davis}, {Ginsburg}, {Price-Whelan}, {Kerzendorf}, {Conley}, {Crighton},
  {Barbary}, {Muna}, {Ferguson}, {Grollier}, {Parikh}, {Nair}, {Unther},
  {Deil}, {Woillez}, {Conseil}, {Kramer}, {Turner}, {Singer}, {Fox}, {Weaver},
  {Zabalza}, {Edwards}, {Azalee Bostroem}, {Burke}, {Casey}, {Crawford},
  {Dencheva}, {Ely}, {Jenness}, {Labrie}, {Lim}, {Pierfederici}, {Pontzen},
  {Ptak}, {Refsdal}, {Servillat}, \& {Streicher}}]{astropy:2013}
{Astropy Collaboration}, {Robitaille}, T.~P., {Tollerud}, E.~J., {et~al.} 2013,
  \aap, 558, A33, \dodoi{10.1051/0004-6361/201322068}

\bibitem[{{Astropy Collaboration} {et~al.}(2018){Astropy Collaboration},
  {Price-Whelan}, {Sip{\H{o}}cz}, {G{\"u}nther}, {Lim}, {Crawford}, {Conseil},
  {Shupe}, {Craig}, {Dencheva}, {Ginsburg}, {Vand erPlas}, {Bradley},
  {P{\'e}rez-Su{\'a}rez}, {de Val-Borro}, {Aldcroft}, {Cruz}, {Robitaille},
  {Tollerud}, {Ardelean}, {Babej}, {Bach}, {Bachetti}, {Bakanov}, {Bamford},
  {Barentsen}, {Barmby}, {Baumbach}, {Berry}, {Biscani}, {Boquien}, {Bostroem},
  {Bouma}, {Brammer}, {Bray}, {Breytenbach}, {Buddelmeijer}, {Burke},
  {Calderone}, {Cano Rodr{\'\i}guez}, {Cara}, {Cardoso}, {Cheedella}, {Copin},
  {Corrales}, {Crichton}, {D'Avella}, {Deil}, {Depagne}, {Dietrich}, {Donath},
  {Droettboom}, {Earl}, {Erben}, {Fabbro}, {Ferreira}, {Finethy}, {Fox},
  {Garrison}, {Gibbons}, {Goldstein}, {Gommers}, {Greco}, {Greenfield},
  {Groener}, {Grollier}, {Hagen}, {Hirst}, {Homeier}, {Horton}, {Hosseinzadeh},
  {Hu}, {Hunkeler}, {Ivezi{\'c}}, {Jain}, {Jenness}, {Kanarek}, {Kendrew},
  {Kern}, {Kerzendorf}, {Khvalko}, {King}, {Kirkby}, {Kulkarni}, {Kumar},
  {Lee}, {Lenz}, {Littlefair}, {Ma}, {Macleod}, {Mastropietro}, {McCully},
  {Montagnac}, {Morris}, {Mueller}, {Mumford}, {Muna}, {Murphy}, {Nelson},
  {Nguyen}, {Ninan}, {N{\"o}the}, {Ogaz}, {Oh}, {Parejko}, {Parley}, {Pascual},
  {Patil}, {Patil}, {Plunkett}, {Prochaska}, {Rastogi}, {Reddy Janga},
  {Sabater}, {Sakurikar}, {Seifert}, {Sherbert}, {Sherwood-Taylor}, {Shih},
  {Sick}, {Silbiger}, {Singanamalla}, {Singer}, {Sladen}, {Sooley},
  {Sornarajah}, {Streicher}, {Teuben}, {Thomas}, {Tremblay}, {Turner},
  {Terr{\'o}n}, {van Kerkwijk}, {de la Vega}, {Watkins}, {Weaver}, {Whitmore},
  {Woillez}, {Zabalza}, \& {Astropy Contributors}}]{astropy:2018}
{Astropy Collaboration}, {Price-Whelan}, A.~M., {Sip{\H{o}}cz}, B.~M., {et~al.}
  2018, \aj, 156, 123, \dodoi{10.3847/1538-3881/aabc4f}

\bibitem[{{Bashi} \& {Zucker}(2019)}]{bashi2019}
{Bashi}, D., \& {Zucker}, S. 2019, \aj, 158, 61,
  \dodoi{10.3847/1538-3881/ab27c9}

\bibitem[{{Bedell} {et~al.}(2014){Bedell}, {Mel{\'e}ndez}, {Bean},
  {Ram{\'\i}rez}, {Leite}, \& {Asplund}}]{bedell2014}
{Bedell}, M., {Mel{\'e}ndez}, J., {Bean}, J.~L., {et~al.} 2014, \apj, 795, 23,
  \dodoi{10.1088/0004-637X/795/1/23}

\bibitem[{{Bedell} {et~al.}(2018){Bedell}, {Bean}, {Mel{\'e}ndez}, {Spina},
  {Ram{\'\i}rez}, {Asplund}, {Alves-Brito}, {dos Santos}, {Dreizler}, {Yong},
  {Monroe}, \& {Casagrande}}]{bedell2018}
{Bedell}, M., {Bean}, J.~L., {Mel{\'e}ndez}, J., {et~al.} 2018, \apj, 865, 68,
  \dodoi{10.3847/1538-4357/aad908}

\bibitem[{{Biazzo} {et~al.}(2015){Biazzo}, {Gratton}, {Desidera}, {Lucatello},
  {Sozzetti}, {Bonomo}, {Damasso}, {Gandolfi}, {Affer}, {Boccato}, {Borsa},
  {Claudi}, {Cosentino}, {Covino}, {Knapic}, {Lanza}, {Maldonado}, {Marzari},
  {Micela}, {Molaro}, {Pagano}, {Pedani}, {Pillitteri}, {Piotto}, {Poretti},
  {Rainer}, {Santos}, {Scandariato}, \& {Zanmar Sanchez}}]{biazzo2015}
{Biazzo}, K., {Gratton}, R., {Desidera}, S., {et~al.} 2015, \aap, 583, A135,
  \dodoi{10.1051/0004-6361/201526375}

\bibitem[{{Booth} \& {Owen}(2020)}]{booth2020}
{Booth}, R.~A., \& {Owen}, J.~E. 2020, \mnras, 493, 5079,
  \dodoi{10.1093/mnras/staa578}

\bibitem[{{Brewer} \& {Fischer}(2017)}]{brewer2017}
{Brewer}, J.~M., \& {Fischer}, D.~A. 2017, \apj, 840, 121,
  \dodoi{10.3847/1538-4357/aa6d53}

\bibitem[{{Bryson} {et~al.}(2021){Bryson}, {Kunimoto}, {Kopparapu}, {Coughlin},
  {Borucki}, {Koch}, {Aguirre}, {Allen}, {Barentsen}, {Batalha}, {Berger},
  {Boss}, {Buchhave}, {Burke}, {Caldwell}, {Campbell}, {Catanzarite},
  {Chandrasekaran}, {Chaplin}, {Christiansen}, {Christensen-Dalsgaard},
  {Ciardi}, {Clarke}, {Cochran}, {Dotson}, {Doyle}, {Duarte}, {Dunham},
  {Dupree}, {Endl}, {Fanson}, {Ford}, {Fujieh}, {Gautier}, {Geary},
  {Gilliland}, {Girouard}, {Gould}, {Haas}, {Henze}, {Holman}, {Howard},
  {Howell}, {Huber}, {Hunter}, {Jenkins}, {Kjeldsen}, {Kolodziejczak},
  {Larson}, {Latham}, {Li}, {Mathur}, {Meibom}, {Middour}, {Morris}, {Morton},
  {Mullally}, {Mullally}, {Pletcher}, {Prsa}, {Quinn}, {Quintana}, {Ragozzine},
  {Ramirez}, {Sanderfer}, {Sasselov}, {Seader}, {Shabram}, {Shporer}, {Smith},
  {Steffen}, {Still}, {Torres}, {Troeltzsch}, {Twicken}, {Uddin}, {Van Cleve},
  {Voss}, {Weiss}, {Welsh}, {Wohler}, \& {Zamudio}}]{bryson2021}
{Bryson}, S., {Kunimoto}, M., {Kopparapu}, R.~K., {et~al.} 2021, \aj, 161, 36,
  \dodoi{10.3847/1538-3881/abc418}

\bibitem[{{Buchhave} \& {Latham}(2015)}]{buchhave2015}
{Buchhave}, L.~A., \& {Latham}, D.~W. 2015, \apj, 808, 187,
  \dodoi{10.1088/0004-637X/808/2/187}

\bibitem[{{Buchhave} {et~al.}(2012){Buchhave}, {Latham}, {Johansen},
  {Bizzarro}, {Torres}, {Rowe}, {Batalha}, {Borucki}, {Brugamyer}, {Caldwell},
  {Bryson}, {Ciardi}, {Cochran}, {Endl}, {Esquerdo}, {Ford}, {Geary},
  {Gilliland}, {Hansen}, {Isaacson}, {Laird}, {Lucas}, {Marcy}, {Morse},
  {Robertson}, {Shporer}, {Stefanik}, {Still}, \& {Quinn}}]{buchhave2012}
{Buchhave}, L.~A., {Latham}, D.~W., {Johansen}, A., {et~al.} 2012, \nat, 486,
  375, \dodoi{10.1038/nature11121}

\bibitem[{{Burke} {et~al.}(2015){Burke}, {Christiansen}, {Mullally}, {Seader},
  {Huber}, {Rowe}, {Coughlin}, {Thompson}, {Catanzarite}, {Clarke}, {Morton},
  {Caldwell}, {Bryson}, {Haas}, {Batalha}, {Jenkins}, {Tenenbaum}, {Twicken},
  {Li}, {Quintana}, {Barclay}, {Henze}, {Borucki}, {Howell}, \&
  {Still}}]{burke2015}
{Burke}, C.~J., {Christiansen}, J.~L., {Mullally}, F., {et~al.} 2015, \apj,
  809, 8, \dodoi{10.1088/0004-637X/809/1/8}

\bibitem[{{Coughlin} {et~al.}(2017){Coughlin}, {Thompson}, \& {Kepler
  Team}}]{coughlin2017}
{Coughlin}, J., {Thompson}, S.~E., \& {Kepler Team}. 2017, in American
  Astronomical Society Meeting Abstracts, Vol. 230, American Astronomical
  Society Meeting Abstracts \#230, 102.04

\bibitem[{{Feeney} {et~al.}(2021){Feeney}, {Wandelt}, \& {Ness}}]{feeney2021}
{Feeney}, S.~M., {Wandelt}, B.~D., \& {Ness}, M.~K. 2021, \mnras, 501, 3258,
  \dodoi{10.1093/mnras/staa3586}

\bibitem[{Fischer \& Valenti(2005)}]{fischer2005}
Fischer, D.~A., \& Valenti, J. 2005, The Astrophysical Journal, 622, 1102,
  \dodoi{10.1086/428383}

\bibitem[{{Fressin} {et~al.}(2013){Fressin}, {Torres}, {Charbonneau}, {Bryson},
  {Christiansen}, {Dressing}, {Jenkins}, {Walkowicz}, \&
  {Batalha}}]{fressin2013}
{Fressin}, F., {Torres}, G., {Charbonneau}, D., {et~al.} 2013, \apj, 766, 81,
  \dodoi{10.1088/0004-637X/766/2/81}

\bibitem[{{Galarza} {et~al.}(2021){Galarza}, {L{\'o}pez-Valdivia},
  {Mel{\'e}ndez}, \& {Lorenzo-Oliveira}}]{galarza2021}
{Galarza}, J.~Y., {L{\'o}pez-Valdivia}, R., {Mel{\'e}ndez}, J., \&
  {Lorenzo-Oliveira}, D. 2021, arXiv e-prints, arXiv:2109.00679.
\newblock \doarXiv{2109.00679}

\bibitem[{{Garc{\'\i}a P{\'e}rez} {et~al.}(2016){Garc{\'\i}a P{\'e}rez},
  {Allende Prieto}, {Holtzman}, {Shetrone}, {M{\'e}sz{\'a}ros}, {Bizyaev},
  {Carrera}, {Cunha}, {Garc{\'\i}a-Hern{\'a}ndez}, {Johnson}, {Majewski},
  {Nidever}, {Schiavon}, {Shane}, {Smith}, {Sobeck}, {Troup}, {Zamora},
  {Weinberg}, {Bovy}, {Eisenstein}, {Feuillet}, {Frinchaboy}, {Hayden},
  {Hearty}, {Nguyen}, {O'Connell}, {Pinsonneault}, {Wilson}, \&
  {Zasowski}}]{garcia2016}
{Garc{\'\i}a P{\'e}rez}, A.~E., {Allende Prieto}, C., {Holtzman}, J.~A.,
  {et~al.} 2016, \aj, 151, 144, \dodoi{10.3847/0004-6256/151/6/144}

\bibitem[{{Ghezzi} {et~al.}(2010){Ghezzi}, {Cunha}, {Smith}, {de Ara{\'u}jo},
  {Schuler}, \& {de la Reza}}]{ghezzi2010}
{Ghezzi}, L., {Cunha}, K., {Smith}, V.~V., {et~al.} 2010, \apj, 720, 1290,
  \dodoi{10.1088/0004-637X/720/2/1290}

\bibitem[{{Ghezzi} {et~al.}(2021){Ghezzi}, {Martinez}, {Wilson}, {Cunha},
  {Smith}, \& {Majewski}}]{ghezzi2021}
{Ghezzi}, L., {Martinez}, C.~F., {Wilson}, R.~F., {et~al.} 2021, \apj, 920, 19,
  \dodoi{10.3847/1538-4357/ac14c3}

\bibitem[{{Ghezzi} {et~al.}(2018){Ghezzi}, {Montet}, \& {Johnson}}]{ghezzi2018}
{Ghezzi}, L., {Montet}, B.~T., \& {Johnson}, J.~A. 2018, \apj, 860, 109,
  \dodoi{10.3847/1538-4357/aac37c}

\bibitem[{{Gonzalez}(1997)}]{gonzalez1997}
{Gonzalez}, G. 1997, \mnras, 285, 403, \dodoi{10.1093/mnras/285.2.403}

\bibitem[{{Griffith} {et~al.}(2021){Griffith}, {Weinberg}, {Johnson}, {Beaton},
  {Garc{\'\i}a-Hern{\'a}ndez}, {Hasselquist}, {Holtzman}, {Johnson},
  {J{\"o}nsson}, {Lane}, {Nataf}, \& {Roman-Lopes}}]{griffith2021}
{Griffith}, E., {Weinberg}, D.~H., {Johnson}, J.~A., {et~al.} 2021, \apj, 909,
  77, \dodoi{10.3847/1538-4357/abd6be}

\bibitem[{Harris {et~al.}(2020)Harris, Millman, van~der Walt, Gommers,
  Virtanen, Cournapeau, Wieser, Taylor, Berg, Smith, Kern, Picus, Hoyer, van
  Kerkwijk, Brett, Haldane, del R{\'{i}}o, Wiebe, Peterson,
  G{\'{e}}rard-Marchant, Sheppard, Reddy, Weckesser, Abbasi, Gohlke, \&
  Oliphant}]{numpy}
Harris, C.~R., Millman, K.~J., van~der Walt, S.~J., {et~al.} 2020, Nature, 585,
  357, \dodoi{10.1038/s41586-020-2649-2}

\bibitem[{Hastie {et~al.}(2001)Hastie, Tibshirani, \& Friedman}]{hastie2001}
Hastie, T., Tibshirani, R., \& Friedman, J. 2001, The Elements of Statistical
  Learning, Springer Series in Statistics (New York, NY, USA: Springer New York
  Inc.)

\bibitem[{{Heiter} \& {Luck}(2003)}]{heiter2003}
{Heiter}, U., \& {Luck}, R.~E. 2003, \aj, 126, 2015, \dodoi{10.1086/378366}

\bibitem[{{Howard} {et~al.}(2012){Howard}, {Marcy}, {Bryson}, {Jenkins},
  {Rowe}, {Batalha}, {Borucki}, {Koch}, {Dunham}, {Gautier}, {Van Cleve},
  {Cochran}, {Latham}, {Lissauer}, {Torres}, {Brown}, {Gilliland}, {Buchhave},
  {Caldwell}, {Christensen-Dalsgaard}, {Ciardi}, {Fressin}, {Haas}, {Howell},
  {Kjeldsen}, {Seager}, {Rogers}, {Sasselov}, {Steffen}, {Basri},
  {Charbonneau}, {Christiansen}, {Clarke}, {Dupree}, {Fabrycky}, {Fischer},
  {Ford}, {Fortney}, {Tarter}, {Girouard}, {Holman}, {Johnson}, {Klaus},
  {Machalek}, {Moorhead}, {Morehead}, {Ragozzine}, {Tenenbaum}, {Twicken},
  {Quinn}, {Isaacson}, {Shporer}, {Lucas}, {Walkowicz}, {Welsh}, {Boss},
  {Devore}, {Gould}, {Smith}, {Morris}, {Prsa}, {Morton}, {Still}, {Thompson},
  {Mullally}, {Endl}, \& {MacQueen}}]{howard2012}
{Howard}, A.~W., {Marcy}, G.~W., {Bryson}, S.~T., {et~al.} 2012, \apjs, 201,
  15, \dodoi{10.1088/0067-0049/201/2/15}

\bibitem[{{Hsu} {et~al.}(2019){Hsu}, {Ford}, {Ragozzine}, \& {Ashby}}]{hsu2019}
{Hsu}, D.~C., {Ford}, E.~B., {Ragozzine}, D., \& {Ashby}, K. 2019, \aj, 158,
  109, \dodoi{10.3847/1538-3881/ab31ab}

\bibitem[{{Hughes} {et~al.}(2007){Hughes}, {Rosner}, \& {Weiss}}]{hughes2007}
{Hughes}, D.~W., {Rosner}, R., \& {Weiss}, N.~O. 2007, {The Solar Tachocline}

\bibitem[{Hunter(2007)}]{matplotlib}
Hunter, J.~D. 2007, Computing in Science \& Engineering, 9, 90,
  \dodoi{10.1109/MCSE.2007.55}

\bibitem[{Ida \& Lin(2004)}]{ida2004}
Ida, S., \& Lin, D. N.~C. 2004, The Astrophysical Journal, 616, 567,
  \dodoi{10.1086/424830}

\bibitem[{{Jofr{\'e}} {et~al.}(2021){Jofr{\'e}}, {Petrucci}, {G{\'o}mez Maqueo
  Chew}, {Ram{\'\i}rez}, {Saffe}, {Martioli}, {Buccino}, {Ma{\v{s}}ek},
  {Garc{\'\i}a}, {Canul}, \& {G{\'o}mez}}]{jofre2021}
{Jofr{\'e}}, E., {Petrucci}, R., {G{\'o}mez Maqueo Chew}, Y., {et~al.} 2021,
  arXiv e-prints, arXiv:2109.04590.
\newblock \doarXiv{2109.04590}

\bibitem[{{Kunimoto} \& {Matthews}(2020)}]{kunimoto2020}
{Kunimoto}, M., \& {Matthews}, J.~M. 2020, \aj, 159, 248,
  \dodoi{10.3847/1538-3881/ab88b0}

\bibitem[{{Liu} {et~al.}(2018){Liu}, {Yong}, {Asplund}, {Feltzing}, {Mustill},
  {Mel{\'e}ndez}, {Ram{\'\i}rez}, \& {Lin}}]{liu2018}
{Liu}, F., {Yong}, D., {Asplund}, M., {et~al.} 2018, \aap, 614, A138,
  \dodoi{10.1051/0004-6361/201832701}

\bibitem[{{Mack} {et~al.}(2014){Mack}, {Schuler}, {Stassun}, \&
  {Norris}}]{mack2014}
{Mack}, Claude~E., I., {Schuler}, S.~C., {Stassun}, K.~G., \& {Norris}, J.
  2014, \apj, 787, 98, \dodoi{10.1088/0004-637X/787/2/98}

\bibitem[{{Maia} {et~al.}(2019){Maia}, {Mel{\'e}ndez}, {Lorenzo-Oliveira},
  {Spina}, \& {Jofr{\'e}}}]{tucci_maia2019}
{Maia}, M.~T., {Mel{\'e}ndez}, J., {Lorenzo-Oliveira}, D., {Spina}, L., \&
  {Jofr{\'e}}, P. 2019, \aap, 628, A126, \dodoi{10.1051/0004-6361/201935952}

\bibitem[{{Maldonado} {et~al.}(2018){Maldonado}, {Villaver}, \&
  {Eiroa}}]{maldonado2018}
{Maldonado}, J., {Villaver}, E., \& {Eiroa}, C. 2018, \aap, 612, A93,
  \dodoi{10.1051/0004-6361/201732001}

\bibitem[{{Maldonado} {et~al.}(2019){Maldonado}, {Villaver}, {Eiroa}, \&
  {Micela}}]{maldonado2019}
{Maldonado}, J., {Villaver}, E., {Eiroa}, C., \& {Micela}, G. 2019, \aap, 624,
  A94, \dodoi{10.1051/0004-6361/201833827}

\bibitem[{{Mayor} {et~al.}(2011){Mayor}, {Marmier}, {Lovis}, {Udry},
  {S{\'e}gransan}, {Pepe}, {Benz}, {Bertaux}, {Bouchy}, {Dumusque}, {Lo Curto},
  {Mordasini}, {Queloz}, \& {Santos}}]{mayor2011}
{Mayor}, M., {Marmier}, M., {Lovis}, C., {et~al.} 2011, arXiv e-prints,
  arXiv:1109.2497.
\newblock \doarXiv{1109.2497}

\bibitem[{{Mel{\'e}ndez} {et~al.}(2009){Mel{\'e}ndez}, {Asplund}, {Gustafsson},
  \& {Yong}}]{melendez2009}
{Mel{\'e}ndez}, J., {Asplund}, M., {Gustafsson}, B., \& {Yong}, D. 2009, \apjl,
  704, L66, \dodoi{10.1088/0004-637X/704/1/L66}

\bibitem[{{Mordasini} {et~al.}(2012){Mordasini}, {Alibert}, {Georgy},
  {Dittkrist}, {Klahr}, \& {Henning}}]{mordasini2012}
{Mordasini}, C., {Alibert}, Y., {Georgy}, C., {et~al.} 2012, \aap, 547, A112,
  \dodoi{10.1051/0004-6361/201118464}

\bibitem[{{Mulders} {et~al.}(2016){Mulders}, {Pascucci}, {Apai}, {Frasca}, \&
  {Molenda-{\.Z}akowicz}}]{mulders2016}
{Mulders}, G.~D., {Pascucci}, I., {Apai}, D., {Frasca}, A., \&
  {Molenda-{\.Z}akowicz}, J. 2016, \aj, 152, 187,
  \dodoi{10.3847/0004-6256/152/6/187}

\bibitem[{{Nagar} {et~al.}(2020){Nagar}, {Spina}, \& {Karakas}}]{nagar2020}
{Nagar}, T., {Spina}, L., \& {Karakas}, A.~I. 2020, \apjl, 888, L9,
  \dodoi{10.3847/2041-8213/ab5dc6}

\bibitem[{{Narang} {et~al.}(2018){Narang}, {Manoj}, {Furlan}, {Mordasini},
  {Henning}, {Mathew}, {Banyal}, \& {Sivarani}}]{narang2018}
{Narang}, M., {Manoj}, P., {Furlan}, E., {et~al.} 2018, \aj, 156, 221,
  \dodoi{10.3847/1538-3881/aae391}

\bibitem[{{Ness} {et~al.}(2022){Ness}, {Wheeler}, {McKinnon}, {Horta}, {Casey},
  {Cunningham}, \& {Price-Whelan}}]{ness2022}
{Ness}, M.~K., {Wheeler}, A.~J., {McKinnon}, K., {et~al.} 2022, \apj, 926, 144,
  \dodoi{10.3847/1538-4357/ac4754}

\bibitem[{{Nibauer} {et~al.}(2021){Nibauer}, {Baxter}, {Jain}, {Van Saders},
  {Beaton}, \& {Teske}}]{nibauer2021}
{Nibauer}, J., {Baxter}, E.~J., {Jain}, B., {et~al.} 2021, \apj, 907, 116,
  \dodoi{10.3847/1538-4357/abd0f1}

\bibitem[{{{\"O}berg} {et~al.}(2011){{\"O}berg}, {Murray-Clay}, \&
  {Bergin}}]{oberg2011}
{{\"O}berg}, K.~I., {Murray-Clay}, R., \& {Bergin}, E.~A. 2011, \apjl, 743,
  L16, \dodoi{10.1088/2041-8205/743/1/L16}

\bibitem[{{Oh} {et~al.}(2018){Oh}, {Price-Whelan}, {Brewer}, {Hogg}, {Spergel},
  \& {Myles}}]{oh2018}
{Oh}, S., {Price-Whelan}, A.~M., {Brewer}, J.~M., {et~al.} 2018, \apj, 854,
  138, \dodoi{10.3847/1538-4357/aaab4d}

\bibitem[{Pedregosa {et~al.}(2011)Pedregosa, Varoquaux, Gramfort, Michel,
  Thirion, Grisel, Blondel, Prettenhofer, Weiss, Dubourg, Vanderplas, Passos,
  Cournapeau, Brucher, Perrot, \& Duchesnay}]{scikit-learn}
Pedregosa, F., Varoquaux, G., Gramfort, A., {et~al.} 2011, Journal of Machine
  Learning Research, 12, 2825

\bibitem[{{Petigura} {et~al.}(2018){Petigura}, {Marcy}, {Winn}, {Weiss},
  {Fulton}, {Howard}, {Sinukoff}, {Isaacson}, {Morton}, \&
  {Johnson}}]{petigura2018}
{Petigura}, E.~A., {Marcy}, G.~W., {Winn}, J.~N., {et~al.} 2018, \aj, 155, 89,
  \dodoi{10.3847/1538-3881/aaa54c}

\bibitem[{{Ram{\'\i}rez} {et~al.}(2019){Ram{\'\i}rez}, {Khanal}, {Lichon},
  {Chanam{\'e}}, {Endl}, {Mel{\'e}ndez}, \& {Lambert}}]{ramirez2019}
{Ram{\'\i}rez}, I., {Khanal}, S., {Lichon}, S.~J., {et~al.} 2019, \mnras, 490,
  2448, \dodoi{10.1093/mnras/stz2709}

\bibitem[{{Ram{\'\i}rez} {et~al.}(2009){Ram{\'\i}rez}, {Mel{\'e}ndez}, \&
  {Asplund}}]{ramirez2009}
{Ram{\'\i}rez}, I., {Mel{\'e}ndez}, J., \& {Asplund}, M. 2009, \aap, 508, L17,
  \dodoi{10.1051/0004-6361/200913038}

\bibitem[{{Ram{\'\i}rez} {et~al.}(2014){Ram{\'\i}rez}, {Mel{\'e}ndez}, \&
  {Asplund}}]{ramirez2014}
---. 2014, \aap, 561, A7, \dodoi{10.1051/0004-6361/201322558}

\bibitem[{{Ram{\'\i}rez} {et~al.}(2011){Ram{\'\i}rez}, {Mel{\'e}ndez},
  {Cornejo}, {Roederer}, \& {Fish}}]{ramirez2011}
{Ram{\'\i}rez}, I., {Mel{\'e}ndez}, J., {Cornejo}, D., {Roederer}, I.~U., \&
  {Fish}, J.~R. 2011, \apj, 740, 76, \dodoi{10.1088/0004-637X/740/2/76}

\bibitem[{{Ram{\'{\i}}rez} {et~al.}(2015){Ram{\'{\i}}rez}, {Khanal}, {Aleo},
  {Sobotka}, {Liu}, {Casagrande}, {Mel{\'e}ndez}, {Yong}, {Lambert}, \&
  {Asplund}}]{ramirez2015}
{Ram{\'{\i}}rez}, I., {Khanal}, S., {Aleo}, P., {et~al.} 2015, \apj, 808, 13,
  \dodoi{10.1088/0004-637X/808/1/13}

\bibitem[{Rice \& Armitage(2003)}]{rice2003}
Rice, W. K.~M., \& Armitage, P.~J. 2003, The Astrophysical Journal, 598, L55,
  \dodoi{10.1086/380390}

\bibitem[{{Saffe} {et~al.}(2016){Saffe}, {Flores}, {Jaque Arancibia},
  {Buccino}, \& {Jofr{\'e}}}]{saffe2016}
{Saffe}, C., {Flores}, M., {Jaque Arancibia}, M., {Buccino}, A., \&
  {Jofr{\'e}}, E. 2016, \aap, 588, A81, \dodoi{10.1051/0004-6361/201528043}

\bibitem[{{Saffe} {et~al.}(2017){Saffe}, {Jofr{\'e}}, {Martioli}, {Flores},
  {Petrucci}, \& {Jaque Arancibia}}]{saffe2017}
{Saffe}, C., {Jofr{\'e}}, E., {Martioli}, E., {et~al.} 2017, \aap, 604, L4,
  \dodoi{10.1051/0004-6361/201731430}

\bibitem[{{Santos} {et~al.}(2004){Santos}, {Israelian}, \&
  {Mayor}}]{santos2004}
{Santos}, N.~C., {Israelian}, G., \& {Mayor}, M. 2004, \aap, 415, 1153,
  \dodoi{10.1051/0004-6361:20034469}

\bibitem[{{Sayeed} {et~al.}(2021){Sayeed}, {Huber}, {Wheeler}, \&
  {Ness}}]{sayeed2021}
{Sayeed}, M., {Huber}, D., {Wheeler}, A., \& {Ness}, M.~K. 2021, \aj, 161, 170,
  \dodoi{10.3847/1538-3881/abdf4c}

\bibitem[{{Schlaufman} \& {Laughlin}(2011)}]{schlaufman2011}
{Schlaufman}, K.~C., \& {Laughlin}, G. 2011, \apj, 738, 177,
  \dodoi{10.1088/0004-637X/738/2/177}

\bibitem[{{Schuler} {et~al.}(2011){Schuler}, {Flateau}, {Cunha}, {King},
  {Ghezzi}, \& {Smith}}]{schuler2011}
{Schuler}, S.~C., {Flateau}, D., {Cunha}, K., {et~al.} 2011, \apj, 732, 55,
  \dodoi{10.1088/0004-637X/732/1/55}

\bibitem[{{Schuler} {et~al.}(2015){Schuler}, {Vaz}, {Katime Santrich}, {Cunha},
  {Smith}, {King}, {Teske}, {Ghezzi}, {Howell}, \& {Isaacson}}]{schuler2015}
{Schuler}, S.~C., {Vaz}, Z.~A., {Katime Santrich}, O.~J., {et~al.} 2015, \apj,
  815, 5, \dodoi{10.1088/0004-637X/815/1/5}

\bibitem[{{Sousa} {et~al.}(2008){Sousa}, {Santos}, {Mayor}, {Udry},
  {Casagrande}, {Israelian}, {Pepe}, {Queloz}, \& {Monteiro}}]{sousa2008}
{Sousa}, S.~G., {Santos}, N.~C., {Mayor}, M., {et~al.} 2008, \aap, 487, 373,
  \dodoi{10.1051/0004-6361:200809698}

\bibitem[{{Sousa} {et~al.}(2019){Sousa}, {Adibekyan}, {Santos}, {Mortier},
  {Barros}, {Delgado-Mena}, {Demangeon}, {Israelian}, {Faria}, {Figueira},
  {Rojas-Ayala}, {Tsantaki}, {Andreasen}, {Brand{\~a}o}, {Ferreira},
  {Montalto}, \& {Santerne}}]{sousa2019}
{Sousa}, S.~G., {Adibekyan}, V., {Santos}, N.~C., {et~al.} 2019, \mnras, 485,
  3981, \dodoi{10.1093/mnras/stz664}

\bibitem[{{Souto} {et~al.}(2019){Souto}, {Allende Prieto}, {Cunha},
  {Pinsonneault}, {Smith}, {Garcia-Dias}, {Bovy}, {Garc{\'\i}a-Hern{\'a}ndez},
  {Holtzman}, {Johnson}, {J{\"o}nsson}, {Majewski}, {Shetrone}, {Sobeck},
  {Zamora}, {Pan}, \& {Nitschelm}}]{souto2019}
{Souto}, D., {Allende Prieto}, C., {Cunha}, K., {et~al.} 2019, \apj, 874, 97,
  \dodoi{10.3847/1538-4357/ab0b43}

\bibitem[{{Teske} {et~al.}(2015){Teske}, {Ghezzi}, {Cunha}, {Smith}, {Schuler},
  \& {Bergemann}}]{teske2015}
{Teske}, J.~K., {Ghezzi}, L., {Cunha}, K., {et~al.} 2015, \apjl, 801, L10,
  \dodoi{10.1088/2041-8205/801/1/L10}

\bibitem[{Teske {et~al.}(2016)Teske, Khanal, \& Ram{\'{\i}}rez}]{teske2016}
Teske, J.~K., Khanal, S., \& Ram{\'{\i}}rez, I. 2016, The Astrophysical
  Journal, 819, 19, \dodoi{10.3847/0004-637x/819/1/19}

\bibitem[{{Ting} \& {Weinberg}(2022)}]{ting2022}
{Ting}, Y.-S., \& {Weinberg}, D.~H. 2022, \apj, 927, 209,
  \dodoi{10.3847/1538-4357/ac5023}

\bibitem[{{Tucci Maia} {et~al.}(2014){Tucci Maia}, {Mel{\'e}ndez}, \&
  {Ram{\'{\i}}rez}}]{tucci_maia2014}
{Tucci Maia}, M., {Mel{\'e}ndez}, J., \& {Ram{\'{\i}}rez}, I. 2014, \apjl, 790,
  L25, \dodoi{10.1088/2041-8205/790/2/L25}

\bibitem[{Virtanen {et~al.}(2020)Virtanen, Gommers, Oliphant, Haberland, Reddy,
  Cournapeau, Burovski, Peterson, Weckesser, Bright, {van der Walt}, Brett,
  Wilson, Millman, Mayorov, Nelson, Jones, Kern, Larson, Carey, Polat, Feng,
  Moore, {VanderPlas}, Laxalde, Perktold, Cimrman, Henriksen, Quintero, Harris,
  Archibald, Ribeiro, Pedregosa, {van Mulbregt}, \& {SciPy 1.0
  Contributors}}]{scipy}
Virtanen, P., Gommers, R., Oliphant, T.~E., {et~al.} 2020, Nature Methods, 17,
  261, \dodoi{10.1038/s41592-019-0686-2}

\bibitem[{{Wang} \& {Fischer}(2015)}]{wang2015}
{Wang}, J., \& {Fischer}, D.~A. 2015, \aj, 149, 14,
  \dodoi{10.1088/0004-6256/149/1/14}

\bibitem[{{Weinberg} {et~al.}(2019){Weinberg}, {Holtzman}, {Hasselquist},
  {Bird}, {Johnson}, {Shetrone}, {Sobeck}, {Allende Prieto}, {Bizyaev},
  {Carrera}, {Cohen}, {Cunha}, {Ebelke}, {Fernandez-Trincado},
  {Garc{\'\i}a-Hern{\'a}ndez}, {Hayes}, {J{\"o}nsson}, {Lane}, {Majewski},
  {Malanushenko}, {M{\'e}sz{\'a}ros}, {Nidever}, {Nitschelm}, {Pan}, {Rix},
  {Rybizki}, {Schiavon}, {Schneider}, {Wilson}, \& {Zamora}}]{weinberg2019}
{Weinberg}, D.~H., {Holtzman}, J.~A., {Hasselquist}, S., {et~al.} 2019, \apj,
  874, 102, \dodoi{10.3847/1538-4357/ab07c7}

\bibitem[{{Weinberg} {et~al.}(2022){Weinberg}, {Holtzman}, {Johnson}, {Hayes},
  {Hasselquist}, {Shetrone}, {Ting}, {Beaton}, {Beers}, {Bird}, {Bizyaev},
  {Blanton}, {Cunha}, {Fern{\'a}ndez-Trincado}, {Frinchaboy},
  {Garc{\'\i}a-Hern{\'a}ndez}, {Griffith}, {Johnson}, {J{\"o}nsson}, {Lane},
  {Leung}, {Mackereth}, {Majewski}, {M{\'e}sz{\'a}ros}, {Nitschelm}, {Pan},
  {Schiavon}, {Schneider}, {Schultheis}, {Smith}, {Sobeck}, {Stassun},
  {Stringfellow}, {Vincenzo}, {Wilson}, \& {Zasowski}}]{weinberg2022}
{Weinberg}, D.~H., {Holtzman}, J.~A., {Johnson}, J.~A., {et~al.} 2022, \apjs,
  260, 32, \dodoi{10.3847/1538-4365/ac6028}

\bibitem[{{W}es {M}c{K}inney(2010)}]{pandas}
{W}es {M}c{K}inney. 2010, in {P}roceedings of the 9th {P}ython in {S}cience
  {C}onference, ed. {S}t\'efan van~der {W}alt \& {J}arrod {M}illman, 56 -- 61,
  \dodoi{10.25080/Majora-92bf1922-00a}

\bibitem[{{Wilson} {et~al.}(2018){Wilson}, {Teske}, {Majewski}, {Cunha},
  {Smith}, {Souto}, {Bender}, {Mahadevan}, {Troup}, {Allende Prieto},
  {Stassun}, {Skrutskie}, {Almeida}, {Garc{\'\i}a-Hern{\'a}ndez}, {Zamora}, \&
  {Brinkmann}}]{wilson2018}
{Wilson}, R.~F., {Teske}, J., {Majewski}, S.~R., {et~al.} 2018, \aj, 155, 68,
  \dodoi{10.3847/1538-3881/aa9f27}

\bibitem[{{Winn} \& {Fabrycky}(2015)}]{winn2015}
{Winn}, J.~N., \& {Fabrycky}, D.~C. 2015, \araa, 53, 409,
  \dodoi{10.1146/annurev-astro-082214-122246}

\bibitem[{{Wittenmyer} {et~al.}(2020){Wittenmyer}, {Wang}, {Horner}, {Butler},
  {Tinney}, {Carter}, {Wright}, {Jones}, {Bailey}, {O'Toole}, \&
  {Johns}}]{wittenmyer2020}
{Wittenmyer}, R.~A., {Wang}, S., {Horner}, J., {et~al.} 2020, \mnras, 492, 377,
  \dodoi{10.1093/mnras/stz3436}

\bibitem[{{Wright} {et~al.}(2012){Wright}, {Marcy}, {Howard}, {Johnson},
  {Morton}, \& {Fischer}}]{wright2012}
{Wright}, J.~T., {Marcy}, G.~W., {Howard}, A.~W., {et~al.} 2012, \apj, 753,
  160, \dodoi{10.1088/0004-637X/753/2/160}

\bibitem[{{Yang} {et~al.}(2020){Yang}, {Xie}, \& {Zhou}}]{yang2020}
{Yang}, J.-Y., {Xie}, J.-W., \& {Zhou}, J.-L. 2020, \aj, 159, 164,
  \dodoi{10.3847/1538-3881/ab7373}

\bibitem[{{Zhu} {et~al.}(2018){Zhu}, {Petrovich}, {Wu}, {Dong}, \&
  {Xie}}]{zhu2018}
{Zhu}, W., {Petrovich}, C., {Wu}, Y., {Dong}, S., \& {Xie}, J. 2018, \apj, 860,
  101, \dodoi{10.3847/1538-4357/aac6d5}

\end{thebibliography}
\bibliographystyle{aasjournal}



\end{document}